\RequirePackage{lineno}
\documentclass[aps,prb,floatfix]{revtex4}

\usepackage{color,graphicx}
\usepackage{epic,epsf}
\usepackage{psboxit}
\usepackage{fancybox}
\usepackage{stmaryrd}
\usepackage{amssymb}
\usepackage{amsmath}
\usepackage{longtable}
\input{amssym.def}

\begin{document}

\title{{\rm Phys.~Rev.~B} 81{\rm, 075319 (2010)}\\[.2cm]Time-evolution of grain size distributions in random nucleation and growth crystallization processes}
\author{Anthony V.~Teran}
\author{Andreas Bill\footnote{abill$@$csulb.edu; author to whom correspondence should be addressed.}}
\affiliation{Department of Physics \& Astronomy, California State University Long Beach, 1250 Bellflower Blvd., Long Beach, California 90840, USA}
\author{Ralf B.~Bergmann}
\affiliation{Institute for Applied Beam Technology (BIAS), Klagenfurter Str.~2, 28359 Bremen, Germany}
\date{Received 1 July, 2009l published 24 February 2010}

\begin{abstract}
We study the time dependence of the grain size distribution $N(\mathbf{r},t)$ during crystallization of a $d-$dimensional solid. A partial differential equation including a source term for nuclei and a growth law for grains is solved analytically for any dimension $d$. We discuss solutions obtained for processes described by the Kolmogorov-Avrami-Mehl-Johnson model for random nucleation and growth (RNG). Nucleation and growth are set on the same footing, which leads to a time-dependent decay of both effective rates. We analyze in detail how model parameters, the dimensionality of the crystallization process, and time influence the shape of the distribution. The calculations show that the dynamics of the effective nucleation {\it and} effective growth rates play an essential role in determining the final form of the distribution obtained at full crystallization. We demonstrate that for one class of nucleation and growth rates the distribution evolves in time into the logarithmic-normal (lognormal) form discussed earlier by Bergmann and Bill [J.~Cryst.~Growth {\bf 310}, 3135 (2008)]. We also obtain an analytical expression for the finite maximal grain size at all times. The theory allows for the description of a variety of RNG crystallization processes in thin films and bulk materials. Expressions useful for experimental data analysis are presented for the grain size distribution and the moments in terms of fundamental and measurable parameters of the model.
\end{abstract}

\maketitle

\section{Introduction}\label{s:introduction}

The micromorphology of solids impacts in an essential way their mechanical, electronic, optical and magnetic properties. It is thus an important task to characterize properly the granularity and homogeneity of materials. This allows in particular the determination or tailoring of their functionality for the development of new microdevices and nanodevices. 
Crystallization and treatment processes define the microstructure of a material. These time-dependent processes occurring under adequate thermodynamic conditions are generally spatially inhomogeneous, generating grains with different sizes, shape and orientation.

One of the main physical observable describing the resulting microstructure is the grain size distribution (GSD), $N(\mathbf{r},t)$, which is a quantitative determination of how many grains of a certain size are found in the sample at a given time.\cite{schmelzer05} The vector $\mathbf{r}$ is some entity modeling the size and shape of the grains ({\it e.g.} diameter of spherical, or semi-axes of ellipsoidal grains, number of atoms in the grains, mass of the grains, etc.) and $t$ is the time. The main purpose of the present work is to determine and analyze the time-evolution of the grain size distribution obtained from the resolution of a partial differential equation (PDE), describing the crystallization of an amorphous solid in $d$ dimensions. The structure of this first-order PDE is motivated by the study of crystallization processes and involves two functions ${\cal D}(\mathbf{r},t)$ and $\mathbf{v}(\mathbf{r},t)$ that describe the source of nuclei and the growth of grains, respectively.

The crystallization of very different materials leads to distribution laws such as the normal, the Weibull, the Gamma (see, {\it e.g.}, Ref.~\onlinecite{distributionlaws}), or most notably the logarithmic-normal (lognormal) distribution.\cite{kumomi03,bergmann98,bergmann97,bergmann97b,bergmann96,kumomi99} The shape of the distribution depends on the mechanisms involved in the formation of polycrystalline materials. When fitting experimental data the choice of the distribution is often not univocal due to the uncertainty of the sampling size considered and the precision of the measurements performed.\cite{redner90} Nevertheless, it is remarkable that the great variety of crystallization processes can be described by only a few distribution laws. This points towards the fact that the detailed knowledge of the interactions and mechanisms involved in the crystallization of solids may determine the shape of grains, but is not required for the determination of the GSD (see Refs.~\onlinecite{schmelzer05} and \onlinecite{shinucleation} and references therein.)

Many theoretical approaches have been proposed to describe the formation of grains during crystallization and they can be divided in two main groups: those that describe the process from an analytical point of view,\cite{gelbard79,shinucleation,sekimoto,axe86,krapivsky,jun05} mainly in one dimension, and those that are based on numerical first principle calculations.\cite{distributionlaws,axe86,crespo,soderlund98,farjas} The latter group heavily relies on numerical approaches applied to statistical physics. We take the first, complementary approach and derive a closed analytical solution to a phenomenological model that contains the main ingredients of crystallization in $d$ dimensions. The PDE and the obtained solution are rather general, but we limit our discussion in this paper to the time-evolution of the grain size distribution for crystallization determined by random nucleation and growth (RNG) processes. This type of crystallization occurs, for example, in amorphous Si thin films.\cite{bergmann97,bergmann97b,kumomi99}

The physical picture underlying the present study is motivated by the Kolmogorov-Avrami-Mehl-Johnson (KAMJ) theory.\cite{kolmogorov37,avrami39,mehljohnson} Starting from a $d-$dimensional solid in the amorphous phase, we assume that nuclei form randomly and homogeneously over the volume of the sample and over time with a constant microscopic rate $I_0$. Each nucleus subsequently grows at constant microscopic rate $v_0$ into a grain until it impinges on other growing grains, at which point the growth in the direction perpendicular to the interface stops. We do not include coalescence of grains in our model. The central result of the KAMJ theory is the analytical determination of the volume fraction of untransformed material $Y(t)$ during crystallization \cite{avrami39} (see Sec.~\ref{ss:RNG}). That the KAMJ result is correct within the assumptions made is well documented,\cite{tomellini,KAMJgeneralization} although it was shown to be the large time limit of a more general expression for the transient process.\cite{shinucleation} Several interesting extensions and modifications of the theory have also been discussed to account for a variety of other crystallization phenomena.\cite{shinucleation,KAMJgeneralization} Here we work within the framework of KAMJ's original model to derive and describe the grain size distribution.

Several groups have studied the grain size distribution during first-order phase transitions taking into account the KAMJ result in their considerations.\cite{sekimoto,axe86,krapivsky,jun05,crespo}  Most notably, Sekimoto considered a partial differential equation in which appears an effective (sometimes also termed actual, transient or average) nucleation rate $I(t) = I_0 Y(t)$ for his study of one-dimensional magnetism.\cite{sekimoto} Similar work followed that was made in the same spirit.\cite{axe86,krapivsky,jun05,crespo} This effective time-dependence is a direct consequence of the time decay of the fraction of available space for nucleation of the new stable phase, derived by KAMJ.\cite{kolmogorov37,avrami39,sekimoto,axe86,shinucleation,krapivsky,jun05,distributionlaws,crespo,soderlund98,farjas} On the other hand, the growth rate has been considered constant, $v(t)=v_0$ in all these models.

\begin{figure}[h]
\includegraphics[width=0.3\textwidth,clip=true]{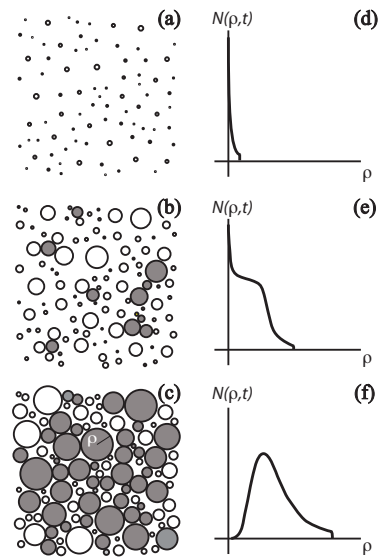}
\unitlength1cm
\caption{\label{fig:sketchvt} Sketch of the nucleation and growth crystallization process for which the theory is developed. See text for a discussion of the figure. Represented are snapshots and distribution $N(\rho,t)$ ($\rho$ is the radius of a grain and $t$ the time) at early stages (a,d), intermediate stages (b,e), and later stages (c,f) of crystallization. The grain size distribution is calculated exactly for all cases in this work. Grey-shaded grains have stopped growing because of impingement, while white grains grow at a rate $v_0$. }
\end{figure}

While staying in the spirit of the KAMJ model by considering constant microscopic nucleation and growth rates $I_0$ and $v_0$ respectively, our approach differs in an essential way with past work on the grain size distribution. Next to the effective time-dependent nucleation rate $I(t) = I_0Y(t)$ first derived by Kolmogorov and Avrami in Refs.~\cite{kolmogorov37,avrami39}, we also introduce an effective time-dependent growth rate $v(t)$. The physical justification of this time dependence not considered previously lies in the fact that the reduced fraction calculated by KAMJ not only reduces the actual nucleation rate, but also affects the average growth rate. This is shown schematically in Fig.~\ref{fig:sketchvt}. The left column displays snapshots of the grain distribution found at early, intermediate and later stages of crystallization. Important for the present discussion are the shaded grains, which cannot grow further due to impingement. At a given time some grains (in white) grow at a constant microscopic rate $v_0$, while others have zero growth rate (shaded grains). In average, the effective growth rate will be less than $v_0$. Comparing Figs.~\ref{fig:sketchvt} (b) and \ref{fig:sketchvt}(c) we note that the number of shaded grains increases with time, implying that the average growth rate decays in time. The right column shows schematically the shape of the grain size distribution $N(\rho,t)$ at these stages of crystallization. The variable $\rho$ is the grain radius. The goal of the paper is to derive analytically and describe, as a function of time and value of parameters, this grain size distribution. All the following figures provide exact results for the sketched GSD of Fig.~\ref{fig:sketchvt}.

The above considerations can be summarized by stating that we place nucleation and growth on the same footing. As will be seen in the following sections, the introduction of time dependence is necessary to describe properly the crystallization of an amorphous solid. Our model and derivation leads to substantial differences with respect to those studied previously. For example, in contrast to previous work our derivation leads to an explicit truncated lognormal-type distribution.\cite{bergmann08} In one dimension it is the lognormal distribution. This result is {\it derived}, not postulated. The time-dependence of the average growth rate is essential to obtain this result. The truncation of the GSD is another feature of our model that follows from the effective time dependent growth rate: we obtain a maximal grain size at all times. Previous analytical works do not contain such a physical cutoff; they decay to zero only in the limit of the infinite grain size. Finally, we obtain an analytical solution of the equations for {\it any} dimension $d$ of the crystal, while previous analytical considerations were limited to one dimension. This is because we are able to calculate the inverse Laplace transform for any dimension $d$.

Using the model just described, this paper provides a detailed derivation and generalization to $d-$dimensions of the remarkable result presented in Ref.~\onlinecite{bergmann08} for three dimensions. We establish classes (subsets) of solutions of our partial differential equation one of which leads to a lognormal distribution in the asymptotic limit of large times. The proposed determination of lognormal-like distributions is interesting in itself because contrary to the usual derivation it does not rely on a probabilistic argument.\cite{kolmogorov41,epstein47,soderlund98,redner90} We discuss in depth the conditions under which such distribution is found in our model, and how it relates to existing derivations of the lognormal distribution in Sec.~\ref{ss:remarks}. We emphasize at this point already that the lognormal asymptotic form is not obtained for the general solution of the equation we establish, but only for a certain class of functions $\mathbf{v}(\mathbf{r},t)$ and ${\cal D}(\mathbf{r},t)$. This class is made of functional forms known to be relevant for the description of random nucleation and growth crystallization processes and involves a time-dependence of the actual nucleation and growth rates.

To end this introduction, we point out that the results provided in the following sections may be relevant to a variety of topics and phenomena. Indeed, the concept of a distribution law is very general and occurs in many fields; for example, the fragmentation of solids,\cite{kolmogorov41,epstein47} gas evaporation,\cite{soderlund98}, the distribution of mass in galaxies,\cite{fontana06} the growth of biological tissues,\cite{limpert01} the distribution of firms as a function of its number of employees,\cite{aitchinson69} the production of scientific publications as a function of the number of researchers in a group,\cite{shockley57} etc.
All these processes are described in terms of a distribution $N(\mathbf{r},t)$, where $\mathbf{r}$ must be appropriately defined and is generally a scalar $r$. In the examples just enumerated, the experimental data can actually all be fitted with more or less accuracy by a lognormal distribution or a composition thereof.\cite{aitchinson69,crow88,limpert01,espiau06} Because of the general structure of the partial differential equation discussed here, and the fact that the lognormal distribution emerges from fairly general principles, it cannot be excluded that the results of the present paper might be applicable to other phenomena in nature, such as those mentioned above.

This paper is structured as follows. In Sec.~\ref{s:model} we describe the partial differential equation for the time-dependent grain size distribution $N(\mathbf{r},t)$ and its physical meaning for RNG crystallization models. In this context we introduce the functional forms for nucleation and growth rates ${\cal D}(\mathbf{r},t)$ and $\mathbf{v}(\mathbf{r},t)$, obtained in the context of the KAMJ model.\cite{kolmogorov37,avrami39,mehljohnson} In Sec.~\ref{s:solutions} we present the general solution of the PDE leading to the central result [Eq.~\eqref{GenSol1} or \eqref{GenSolrt}]. We also calculate the time-dependence of the maximal grain size [Eq.~\eqref{gammax} or \eqref{gammaxrt}] and provide a simple relation between basic measurable parameters of the model [Eq.~\eqref{paramrelation} or \eqref{ginf}]. The general solution is divided into classes according to the time dependence of the nucleation and growth rates and the dimensionality of the crystallization process. We show that for one specific class, the distribution evolves into a lognormal form in the asymptotic limit $t\to\infty$.\cite{bergmann08} We elaborate on the relevance of this result in the following section. In Sec.~\ref{s:timeevolution}, we present a detailed analysis of the time- and model-parameter dependences of the GSD and discuss our results. We also provide in Sec.~\ref{ss:remarks} a discussion of our model in the context of other PDEs and derivations of the lognormal distribution. Finally, we conclude with Sec.~\ref{s:conclusions}.  An example of application of the theory to the analysis of experimental data can be found in Refs.~\onlinecite{bergmann08} and \onlinecite{billMRS09}, where the case of solid phase crystallization of amorphous silicon thin films has been considered.

The theoretical calculations are performed with dimensionless quantities. Since one of the purposes of this paper is to provide simple closed analytical forms of the time-dependent distribution that can be used to analyze data and extract fundamental parameters of a physical system we added two appendixes. Appendix \ref{a:parameters} gives a convenient table of all parameters, variables and relations between constants defined in the paper, and Appendix Sec.~\ref{a:solutioninrandt} summarizes the main results written in quantities with physical dimensions.

\section{Model}\label{s:model}

The theory introduced in this section is potentially applicable to a variety of phenomena in nature. Therefore, we first describe in part (Sec.~\ref{ss:diffeq}) the differential equation for the distribution $N(\mathbf{r},t)$ in general terms. We then write in Sec.~\ref{ss:diffeqRNG} the equation specifically for RNG crystallization.

\subsection{Differential equation for the distribution $N(\mathbf{r},t)$}\label{ss:diffeq}

We are interested in phenomena describable by a time-dependent distribution $N(\mathbf{r},t)$ of a certain entity defined in terms of a $d-$dimensional vector $\mathbf{r}$. In the case of crystallization processes, the entity is the crystalline grain and the vector $\mathbf{r}$ describes the shape and size of a grain. For example, if the entities are grains of ellipsoidal shape, they may be described by the semi-axes $r_j$ ($j=1,\dots,d$) and $\mathbf{r} = (r_1,\dots,r_d)$. Another possible choice of $\mathbf{r}$ is to describe the grain in terms of the number of constituting atoms or the mass $M$, in which case $r=M$ is a scalar (as discussed in the next section, the latter choice leads to a non-linear differential equation, which cannot be solved analytically, while the first choice leads to a set of linear differential equations that can be solved analytically). These two examples show that, in general, the dimension $d$ of the vector $\mathbf{r}$ may be different from the spatial dimension. In the considerations of the present paper the two dimensions will turn out to be identical. Note that this definition of $N(\mathbf{r},t)$ presupposes that the distribution is spatially homogeneous and does, therefore, not depend on the position vector $\mathbf{x}$ in the sample.

We assume that the dynamics of the distribution is affected by two processes. One is a source-and-sink term ${\cal D}(\mathbf{r},t)$ that describes the creation and/or annihilation of the entities. The other describes the dynamic growth and/or shrinking of the entities and is defined by the vectorial quantity $\mathbf{v}(\mathbf{r},t)$. The analytical calculations of this paper can be extended to PDEs that include further terms to account, for example, for the possible coalescence of entities, as long as the equation remains first order and linear. Beyond that, numerical techniques are likely the path to follow, except in one dimension, as shown in Refs.~\onlinecite{sekimoto,axe86,krapivsky}.

The contribution of the source-and-sink term to the time-evolution of the grain size distribution $N(\mathbf{r},t)$ is given by
\begin{eqnarray}
\frac{\partial N(\mathbf{r},t)}{\partial t} = {\cal D}(\mathbf{r},t).
\end{eqnarray}
In the following we consider the separation of variables ${\cal D}(\mathbf{r},t) = I(t)D(\mathbf{r})$. The function $D(\mathbf{r})$ may often be modeled by a Gaussian centered on some characteristic quantity $\mathbf{r}_c$ (for example, the radius of a spherical nucleus $\rho_c$), which may be replaced by a Dirac delta distribution in the limit of zero variance.
The time-dependent part of the nucleation rate, $I(t)$, will be discussed in more details in Sec.~\ref{ss:RNG} below.

The growth of a fixed number of entities is described in terms of a general continuity equation in the space of the vector $\mathbf{r}$
\begin{eqnarray}\label{continuitygen}
\frac{\partial N(\mathbf{r},t)}{\partial t} + \mathbf{\nabla}_{\mathbf{r}}\cdot\left[ N(\mathbf{r},t)\,\mathbf{v}(\mathbf{r},t) \right] = 0.
\end{eqnarray}
The growth rate of the grains, $\mathbf{v}(\mathbf{r},t)$, will also be discussed in more details in Sec.~\ref{ss:RNG}.

Assuming that these are the two only contributions to the rate of change in the distribution $N(\mathbf{r},t)$, we are led to the following partial differential equation
\begin{eqnarray}\label{PDEgeneral}
\frac{\partial N(\mathbf{r},t)}{\partial t} + \mathbf{\nabla}_{\mathbf{r}}\cdot\left[ N(\mathbf{r},t)\,\mathbf{v}(\mathbf{r},t) \right] = {\cal D}(\mathbf{r},t).
\end{eqnarray}
We emphasize the point made above that $\mathbf{r}$ is not in general related to a spatial coordinate, but characterizes the entities that are created and grow in time.

The PDE introduced here on physical grounds is well known and displays a large variety of more or less complicated solutions, depending on the ingredients introduced to account for a particular phenomenon. The study of this equation leads to an interesting and unexpected outcome when applied to crystallization processes, as shown in later sections (see also Ref.~\onlinecite{bergmann08}.) A formal general solution can be found for this first-order linear partial differential equation and specific solutions can be obtained by quadrature for well-behaved source-and-sink terms ${\cal D}(\mathbf{r},t)$ and growth rates $\mathbf{v}(\mathbf{r},t)$.\cite{PDEbooks} We discuss in the next sections {\it closed} analytical solutions for growth rates and source terms applicable to random nucleation and growth processes during crystallization of amorphous solids. Note that altered forms of the above equation, some pertinent to other physical situations, have been discussed in the literature (see, {\it e.g.}, Refs.~\onlinecite{gelbard79,brock72,williams91,sekimoto}, \onlinecite{axe86}, and \onlinecite{krapivsky} and references therein.) The choice of $\mathbf{v}(\mathbf{r},t)$ and $\mathcal{D}(\mathbf{r},t)$, the formalism, and the resulting analytical solutions distinguish the present work from others.

\subsection{Differential equation for RNG processes}\label{ss:diffeqRNG}

To apply the differential equation of the previous section to RNG processes, we need to specify the rates ${\cal D}(\mathbf{r},t)$ and $\mathbf{v}(\mathbf{r},t)$. Transmission electron microscope\cite{bergmann97,bergmann97b,bergmann98,kumomi95,espiau06} or atomic force microscope \cite{AFM} measurements performed during crystallization of many different materials reveal grains with a great variety of shapes and sizes. These grains are generally characterized by a scalar quantity, such as the number of atoms, the volume or the mass of the grain. In the present paper, we take an alternative approach, and describe the grains as ellipsoids with semi-axes $r_j$ with $j=1,\dots, d$. For thin films $d=2$, whereas for bulk solids $d=3$. The components of the vector $\mathbf{r}$ defined in the previous section are now given by the semi-axes of the ellipsoid $\mathbf{r} = (r_1,\cdots,r_d)$ with $r_1 \geq \dots \geq r_d \geq 0$. In this representation the dimensionality $d$ of the vector $\mathbf{r}$ coincides with the spatial dimension of the nucleation and growth process.

In general, the experimental representation of the grain size distribution is done in terms of the average radius of a $d-$dimensional sphere, the volume of which equals the measured volume of the grain. For this reason, we later consider the special case of spherical grains to allow comparison of the theory with existing experimental data.\cite{bergmann08,billMRS09} This reduction to spheres in $d$ dimensions has the additional advantage of expressing the grain size distribution in terms of two variables only (the radius of the grain $\rho$ and time $t$) and allows a better understanding of the behavior of the distribution as a function of nucleation and growth rates without the additional difficulty related to the anisotropic growth and orientation of the grains.

In our RNG model, the source ${\cal D}(\mathbf{r},t)$ describes the formation of nuclei (we do not consider the dissolution and coarsening of grains) that is determined by microscopic interactions and the thermodynamical conditions, under which crystallization of a specific material occurs. We assume that nuclei are formed with a critical volume $\Omega_{c,d}$ at a rate $I(t)$. The source term thus takes the form
\begin{eqnarray}\label{Ddelta}
{\cal D}(\mathbf{r}) = I(t)\,D(\mathbf{r}) = I(t)\,\delta\left(\Omega_d - \Omega_{c,d}\right),
\end{eqnarray}
where
\begin{eqnarray}\label{Omegad}
\Omega_d = \omega_d\, r_1\dots r_d,
\quad
\mbox{with}\quad  \omega_d =  \frac{\pi^{d/2}}{\textstyle \Gamma\left[\frac{d}{2}+1\right]},
\end{eqnarray}
gives the volume of a $d-$dimensional ellipsoid, and $\Gamma$ is the gamma function. $\Omega_{c,d}$ is defined as in Eq.~\eqref{Omegad} and is the critical volume of a nucleus, that is, the volume of the smallest grain that can be found in the sample. This simplified expression for the nucleation term \eqref{Ddelta}, which states that all nuclei are formed with the same volume $\Omega_{c,d}$, is consistent with the fact that we do not discuss the details of the nucleation process. It is a good first approximation, except for describing early stages of crystallization.\cite{shinucleation} This will become apparent in Sec.~\ref{s:timeevolution} and the application of the theory presented in Ref.~\onlinecite{billMRS09}.

Once a nucleus is formed, crystallization leads to its growth into a grain. One simplification relevant to some RNG processes is the case where the variation in the growth rate $\mathbf{v}(\mathbf{r},t)$ only weakly depends on $\mathbf{r}$. In this case, Eq.~\eqref{PDEgeneral} can be approximated by
\begin{eqnarray}\label{PDEgeneral1}
\frac{\partial N(\mathbf{r},t)}{\partial t} + \mathbf{v}(t)\cdot \mathbf{\nabla}_{\mathbf{r}} N(\mathbf{r},t) = {\cal D}(\mathbf{r},t) .
\end{eqnarray}
In fact, this equation applies to the case of the solid phase crystallization processes considered in Refs.~\onlinecite{bergmann08} and \onlinecite{billMRS09}. In this case, the growth rate $\mathbf{v}(\mathbf{r},t)$ is independent of $\mathbf{r}$. This non-trivial simplification is one reason for choosing the present formalism in the vector space of $\mathbf{r}$ rather than writing the grain size distribution $N(y,t)$ in terms of a scalar $y$ such as the volume or the number of atoms of a grain. Indeed, in processes such as solid-phase crystallization,\cite{bergmann97,bergmann97b,kumomi99,bergmann08,billMRS09,bergmann98} atoms are present in the immediate vicinity of nuclei and grains at all time. Thus, their contribution to the growth of grains does not require diffusion processes and the rate $\mathbf{v}$ does not depend on the size of the grains.\cite{kolmogorov37,avrami39,mehljohnson} It is essential to realize that writing the equation for random nucleation and growth, in terms of a scalar, such as the number of atoms in the grain, would result in a non-linear dependence of the growth rate on that scalar, which is generally untractable analytically. The present formalism allows to circumvent this problem at the cost of a multidimensional PDE, which turns out to be solvable analytically.

As mentioned above, applications\cite{bergmann08,billMRS09} assume grains of spherical shape and the grain size distribution depends only on the radius $\rho$ of the $d-$dimensional sphere. In the space of semi-axes of the ellipsoid, the vector for a sphere is $\mathbf{r}=(r_1,\dots,r_d) = (\rho,\dots,\rho)$, and the magnitude of the vector appearing in Eqs.~\eqref{Ddelta}-\eqref{PDEgeneral1} is $r = \sqrt{d}\, \rho$. Introducing polar coordinates, where $\rho$ is the radial coordinate, and using the expression for the source term \eqref{Ddelta}, we obtain
\begin{eqnarray}\label{PDEradius}
\frac{\partial N(\rho,t)}{\partial t} + \frac{v(t)}{\rho^{d-1}} \frac{\partial }{\partial \rho}\left[\rho^{d-1} N(\rho,t) \right] = \frac{I(t)}{A_{c,d}}\,\delta(\rho-\rho_c),
\end{eqnarray}
where $A_{c,d} = \omega_d\,d\,\rho_c^{d-1}$ is the surface of the hyperspherical nucleus of radius $\rho_c$ in $d-$dimension.

We derive in the next section the exact solution of the above equation for rates relevant to RNG processes, with the boundary conditions
\begin{subequations}\label{genboudcond}
\begin{eqnarray}
{N}(\rho \leq \rho_c,t) = 0,\quad {N}(\rho\to \infty,t) = 0,\quad  \mbox{for} \quad t\geq t_0,
\end{eqnarray}
and the initial condition
\begin{eqnarray}
{N}(\rho,t) =0, \quad \mbox{for} \quad t\leq t_0,
\end{eqnarray}
\end{subequations}
where $t_0$ is the incubation time. We emphasize that contrary to many approaches, the present formalism does not assume prior knowledge of an initial distribution; there are no grains in the samples until the incubation time is reached.

\subsection{Random nucleation and growth rates}\label{ss:RNG}

We define in this section the effective nucleation and growth rates $I(t)$ and $v(t)$ that appear in Eq.~\eqref{PDEgeneral}. The thermodynamics of nucleation and early stages of crystallization is involved and a subject of its own.\cite{shinucleation,christian} We assume that the conditions necessary for the creation of nuclei and their subsequent growth into grains are fulfilled. The creation of nuclei is described in terms of the existence of a critical average volume $\Omega_c$ and an effective rate $I(t)$ introduced in Eq.~\eqref{Ddelta}. We are mainly interested in the time dependence of the grain size distribution, resulting from RNG processes and consider situations where no coarsening occurs.\cite{bergmann97,bergmann97b,kumomi99} Full crystallization has been completed once each atom of the sample is assigned to a grain. This is a realistic scenario, for example, in solid-phase crystallization of an amorphous sample,\cite{bergmann08,billMRS09} and is the one considered in the KAMJ theory for RNG processes.\cite{bergmann08,kolmogorov37,avrami39,mehljohnson}

The model assumes homogeneous nucleation; nuclei appear randomly in the sample (no pre-existing nuclei or nucleation centers) with a constant microscopic rate $I_0$, irrespective of the presence of already transformed material. To satisfy this condition, the concept of phantom nuclei was introduced  in Ref.~\onlinecite{avrami39}. Avrami obtained an expression relating the effective to the extended volume fraction of transformed material that corrected for the presence of these phantom nuclei and lead him to derive an expression for the fraction of material available at time $t$ for further nucleation and growth\cite{kolmogorov37,avrami39,mehljohnson}
\begin{eqnarray}\label{KAMJ}
Y_n(t) = \exp\left\{- \left(\frac{t-t_0}{t_c}\right)^{n+1}\right\}\,\Theta\left(\frac{t-t_0}{t_c}\right).
\end{eqnarray}
$t_0$ is the incubation time, $t_c$ is the critical crystallization time, and the integer $n$ determines the time dependence of the crystallization process.  $\Theta(t)$ is the Heaviside function.

Later work on grain formation in various systems to which the KAMJ applies ({\it e.g.} Refs.~\onlinecite{sekimoto,axe86,shinucleation,krapivsky,jun05,crespo,soderlund98,farjas} and \onlinecite{christian}) has recognized that though the microscopic nucleation rate $I_0$ is constant, the actual (effective) nucleation rate relevant for the grain size distribution decays in time as $I(t) = I_0Y_n(t)$ because of the reduced fraction of available space for nucleation. This effective rate has been confirmed, refined, and extended to include a variety of nucleation phenomena. For example, the expression for $I(t)$ has been generalized to account more precisely for the early stages of nucleation.\cite{shinucleation} The concept of "phantom nuclei" introduced by Avrami\cite{avrami39} and the importance of which has been thoroughly discussed\cite{tomellini} has been linked to the concept of spatial correlations of nucleation.\cite{sekimoto,tomellini} Further extensions of the model have been proposed to include simultaneous nucleation, non-random processes, coalescence, or to take into account the symmetry of the crystal structure in the transformed phase.\cite{sekimoto,krapivsky, KAMJgeneralization,tomellini}

In the present paper we account for the insight gained in earlier work, but extend the analysis in an essential way by also considering the effective change in the growth rate. This aspect of the growth of grains during crystallization has not been addressed in any previous work. We start with the random and homogeneous nucleation of grains in the sample. As the crystallization process develops, each nucleus grows into a grain at a rate $v_0$. The growth of each grain eventually comes to a halt, either because it is inhibited by the presence of another grain boundary ({\it e.g.}, by impingement) or because crystallization has been completed (no further free atoms are available). Thus, in the same spirit as for the analysis of the effective nucleation rate, the growth rate actually decreases with time (see Fig.~\ref{fig:sketchvt}). The proper account of this effective time-dependent growth rate $v(t)$ turns out to modify, in important ways, the grain size distribution and leads to the lognormal-like form of the distributions observed in experiment.\cite{bergmann08,billMRS09}

Based on Eq.~\eqref{KAMJ}, we thus develop further the theory of RNG by introducing different rates $I(t)$ and $v(t)$ for nucleation and growth, respectively. We postulate that both rates take similar functional form, except, for different critical times $t_{cI}$ and $t_{cv}$ and different power laws in the exponential. As a result, we introduce the following expressions for nucleation and growth rates:
\begin{subequations}\label{rates}
\begin{eqnarray}
\label{It}
I_n(t) &=& I_0 Y_n^I(t) = I_0 \exp\left\{- \left(\frac{t-t_0}{t_{cI}}\right)^{n+1}\right\}\,\Theta\left(\frac{t-t_0}{t_{cI}}\right),\\
\label{vt}
v_m(t) &=& v_0 Y_m^v(t) = v_0 \exp\left\{- \left(\frac{t-t_0}{t_{cv}}\right)^{m+1}\right\}\,\Theta\left(\frac{t-t_0}{t_{cv}}\right).
\end{eqnarray}
\end{subequations}
While the effective nucleation rate [Eq.~\eqref{It}] has been used in previous work\cite{sekimoto,axe86,krapivsky, jun05,crespo,soderlund98,farjas} to determine the grain size distribution, the introduction of the effective growth rate [Eq.~\eqref{vt}] is new. The analytical form of $v_m(t)$ and the choice $m=0$ (exponential time decay) will be justified {\it a posteriori} by analyzing the analytical results and comparing the theory with experimental data.\cite{billMRS09}
It is important to realize that the KAMJ model relies on the fundamental assumption of  constant and homogeneous microscopic nucleation and growth rates, denoted $I_0$ and $v_0$. By introducing the above effective rates [Eqs.~\eqref{rates}], we thus presuppose that the physical picture underlying the present theory is the same as that of the KAMJ model, but we account for the effective time decay of the nucleation and growth rates.

The nucleation and growth rates contain five parameters: $I_0$, $v_0$, $t_0$, $t_{cI}$ and $t_{cv}$. However, if one shifts the time variable by $t_0$, it turns out that only two parameters actually affect the normalized distribution, namely, $v_0$ and the ratio $t_{cv}/t_{cI}$. In addition, the integers $n$ and $m$ determining the power law of the time decay have to be specified. For $m=0$ (or $n=0$), the decay is exponential, and for $m=1$ it is Gaussian. For $m>1$ it is super Gaussian. For $m=-1$, the growth rate is constant, $v_{-1}(t) = e v_0$ and the PDE for $d=1$ reduces to that of Refs.~\onlinecite{sekimoto} and \onlinecite{krapivsky} in the absence of coalescence.
The nucleation rate $I(t)$ is essentially given by the fraction of material available for crystallization and thus $n=d$, the dimensionality of the grain.\cite{avrami39} Consequently, our $t_{cI}$ is the critical time $t_c$ in the conventional KAMJ model. On the other hand, the growth rate $v(t)$ is determined by another mechanism, namely, the inhibition of the grain growth by their neighbors. We therefore expect the exponent $m$ to differ from $n$. Of importance for the following are the values $m=0,1$ and $n=1,2,3$. The case $m=-1$ and $n=1$ has been studied analytically,\cite{sekimoto,krapivsky} taking into account coalescence, which is absent in the present model. Note finally that $m$ must not necessary be an integer, but for simplicity we consider only this case here.

\section{Solution for the extended Kolmogorov-Avrami-Mehl-Johnson model}\label{s:solutions}

We calculate the grain size distribution $N(\mathbf{r},t)$ for the random nucleation and growth crystallization model described in the previous section by solving Eq.~\eqref{PDEgeneral} with Eqs.~\eqref{rates}. This can be achieved formally by various methods as, for example, the Laplace transform or the methods of characteristics.\cite{PDEbooks} An explicit expression for $N(\mathbf{r},t)$ is then obtained by quadrature once $\mathbf{v}(\mathbf{r},t)$ and ${\cal D}(\mathbf{r},t)$ are defined.
For the nucleation and growth rates defined in Eqs.~\eqref{rates}, it is more natural and instructive to use the Laplace transform. Remarkably, we can obtain the inverse Laplace transform in any dimension within the present model. The details of the derivation of $N(\mathbf{r},t)$ are presented in Appendix \ref{a:solution}. We also determine the time-dependent maximal size of the grains that can be found in the sample, and the moments of the distribution.

\subsection{Partial differential equation in dimensionless quantities}\label{ss:dimless}

It is useful to introduce the following dimensionless variables:
\begin{subequations}\label{dimless}
\begin{eqnarray}\label{dimlessvar}
\gamma &=& \frac{\rho}{\rho_m^\infty},\quad \tau = \frac{t}{\sqrt{t_{cv}t_{cI}}},
\end{eqnarray}
and constants
\begin{eqnarray}\label{dimlesscte}
\tau_0 &=& \frac{t_0}{\sqrt{t_{cv}t_{cI}}},\quad t_r = \sqrt{\frac{t_{cv}}{t_{cI}}},\quad
{\cal I}_0 = t_{cI}I_0\rho_m^\infty,\quad {\cal V}_0 = \frac{t_{cv}v_0}{\rho_m^\infty}.
\end{eqnarray}
\end{subequations}
It is important to bear in mind that a simplified notation is used in Eqs.~\eqref{dimless} and the following to avoid overloading expressions with indexes. The dimensionless variable $\gamma$ actually depends on $m$ through $\rho_m^\infty$. This choice of dimensionless radius is made because, in general, $\rho_c\ll \rho_m^\infty$ and the latter is easily measured experimentally while the former is not. Since we consider solutions obtained for different growth laws (distinguished by $m$) separately, no confusion can arise from this simplified notation. In Sec.~\ref{s:timeevolution}, we analyze the dependence of the general grain size distribution on the parameters of the model, that is, for various values of the constants defined in Eq.~\eqref{dimlesscte}.

To solve the partial differential equation \eqref{PDEradius} it is useful to write it in dimensionless variables \eqref{dimless} and to introduce the auxiliary dimensionless function $\tilde{N}(\gamma,\tau) = (\rho_m^{\infty})^2\,A_{\infty,d}\,\gamma^{d-1}N(\gamma,\tau)$, where $A_{\infty,d} = \omega_d\, d\, (\rho_m^{\infty})^{d-1}$  is the surface of the largest hyperspherical grain of radius $\rho_m^\infty$ found in a $d-$dimensional sample at full crystallization. Then, the differential equation.~\eqref{PDEgeneral} or Eq.~\eqref{PDEradius} takes the form
\begin{eqnarray}\label{PDENtildedimless}
\frac{\partial \tilde{N}(\gamma,\tau)}{\partial \tau} + \frac{v_m(\tau)}{t_r}\frac{\partial \tilde{N}(\gamma,\tau)}{\partial \gamma} = t_r \,I_n(\tau)\,\delta(\gamma - \gamma_c),
\end{eqnarray}
with $\gamma_c = \rho_c/\rho_m^\infty$, and the dimensionless nucleation and growth rates
\begin{subequations}\label{ratesdimless}
\begin{eqnarray}
\label{Idimless}
I_n(\tau) &=& {\cal I}_0\, Y_n^I(\tau) = {\cal I}_0\, \exp\left\{-t_r^{n+1}\left(\tau-\tau_0\right)^{n+1}\right\} \Theta\left(\tau-\tau_0\right),\\
\label{vdimless}
v_m(\tau) &=&
{\cal V}_0 Y_m^v(\tau) = {\cal V}_0 \exp\left\{-\left(\frac{\tau-\tau_0}{t_r}\right)^{m+1}\right\} \Theta\left(\tau-\tau_0\right),
\end{eqnarray}
\end{subequations}
It is important to remember that the grain size distribution is given by $N(\gamma,\tau)$, not by $\tilde{N}(\gamma,\tau)$. The latter is only an auxiliary function to bring the partial differential equation in a form that can be solved analytically.

\subsection{Solution of the partial differential equation \eqref{PDENtildedimless}}

We derive in Appendix \ref{a:solution} the solution of Eq.~\eqref{PDEgeneral1}, written in dimensionless quantities as Eq.~\eqref{PDENtildedimless}, for the KAMJ model [Eq.~\eqref{KAMJ}], and for effective nucleation and growth rates [Eqs.~\eqref{ratesdimless}]. We obtain the following grain size distribution for any dimension $d$
\begin{eqnarray}\label{GenSol}
N(\gamma,\tau) &=& \frac{\tilde{N}(\gamma ,\tau)}{(\rho_m^\infty)^2 A_{\infty,d}\,\gamma^{d-1}} \\
&=&
\frac{C_d}{\gamma^{d-1}} \sum_i \exp\left[\left(\frac{\sigma_{m,i}-\tau_0}{t_r}\right)^{m+1} - t_r^{n+1}\left(\sigma_{m,i}-\tau_0\right)^{n+1}\right]\,\Theta(\tau-\sigma_{m,i})\,\Theta\left(\sigma_{m,i}-\tau_0\right),\nonumber
\end{eqnarray}
with the constant prefactor $C_d$ defined by
\begin{eqnarray}\label{Cd}
C_d = \left(\frac{I_0}{v_0}\right) \frac{1}{A_{\infty,d}},
\end{eqnarray}
with $A_{\infty,d}$ defined above Eq.~\eqref{PDENtildedimless}.

The general solution is directly proportional to the ratio $I_0/v_0$ and inversely proportional to $\gamma^{d-1}$. The coefficient $C_d \sim I_0/v_0$ cancels out when considering normalized distributions. Such is the case in the numerical calculations of the next sections and often also in the description of experimental data.\cite{billMRS09}
The functions $\sigma_{m,i}(\gamma,\tau)$ ($i=1,2,\dots$) are solutions of (see appendix \ref{a:solution})
\begin{eqnarray}\label{eqsigmami}
\gamma = \gamma_c + u_m(\sigma_{m,i},\tau),
\end{eqnarray}
with $u_m$ defined by
\begin{eqnarray}\label{umtau}
u_m(a,b)  \equiv \frac{1}{t_r} \int_{a}^bv_m(\tau')\,d\tau'.
\end{eqnarray}
The function $\sigma_{m,i}$ only depends on the growth rate, not on the nucleation rate. For the KAMJ growth rate \eqref{vdimless}, the latter equation has a single solution; the sum over $i$ in Eq.~\eqref{GenSol} and the index $i$ in $\sigma_{m,i}$ can therefore be dropped. Equation \eqref{eqsigmami} can be rewritten in the form as
\begin{eqnarray}\label{sigmagen}
\Gamma\left[\frac{1}{m+1},\left(\frac{\sigma_m - \tau_0}{t_r}\right)^{m+1}\right] = (m+1)\frac{\gamma-\gamma_c}{{\cal V}_0} + \Gamma\left[\frac{1}{m+1},\left(\frac{\tau - \tau_0}{t_r}\right)^{m+1}\right]
\end{eqnarray}
for $m\geq 0$, where $\Gamma[a,x]$ is the upper incomplete gamma function. This equation yields for $m=0$
\begin{subequations}\label{sigmam}
\begin{eqnarray}\label{sigma0}
\sigma_0(\gamma,\tau) = \tau_0 + t_r\,\ln\left( \frac{\gamma-\gamma_c}{{\cal V}_0} + e^{-(\tau-\tau_0)/t_r} \right)^{-1},
\end{eqnarray}
and for $m=1$
\begin{eqnarray}\label{sigma1}
\sigma_1(\gamma,\tau) = \tau_0 + t_r\, \mbox{erf}^{-1}\left[\mbox{erf}\left(\frac{\tau-\tau_0}{t_r}\right) -  \frac{2}{\sqrt{\pi}}\frac{\gamma-\gamma_c}{{\cal V}_0}\right],
\end{eqnarray}
where erf is the error function. For higher values of $m$, Eq.~\eqref{eqsigmami} is implicit and solving it involves special functions.
\end{subequations}

To bring to the fore the physics contained in our general solution for the KAMJ model, we transform the time-dependent product of Heaviside functions appearing in Eq.~\eqref{GenSol} into a size-dependent difference of these functions (see Appendix \ref{a:solution})
\begin{subequations}\label{prodHeavisidefinal}
\begin{eqnarray}
\Theta\left[\tau-\sigma_m(\gamma,\tau)\right]\,\Theta\left[\sigma_m(\gamma,\tau)-\tau_0\right]
&=&
\Theta\left(\gamma-\gamma_c\right) - \Theta\left[\gamma-\gamma_m^{\max} (\tau)\right],\\
\lim_{\tau\to\infty}\,\Theta \left[\tau-\sigma_m(\gamma,\tau)\right]\,\Theta \left[\sigma_m(\gamma,\tau)-\tau_0 \right] &=& \Theta\left(\gamma-\gamma_c\right) - \Theta\left(\gamma-1\right),
\end{eqnarray}
\end{subequations}
where $\gamma_m^{\max}(\tau)$ is determined below. This difference expresses the fact that at all times, only grains with radius between the minimum $\rho_c$ ($\gamma = \gamma_c$) and the maximal size $\rho_m^{\max}(t)$ ($\gamma = \gamma_m^{\max}(\tau)$) can be found in the sample. Thus, Eq.~\eqref{GenSol} now reads as
\begin{eqnarray}\label{GenSol1}
N(\gamma ,\tau) &=&
\frac{C_d}{\gamma^{d-1}} \exp\left[\left(\frac{\sigma_{m}-\tau_0}{t_r}\right)^{m+1} - t_r^{n+1}\left(\sigma_{m}-\tau_0\right)^{n+1}\right] \nonumber\\
&&\times \Big\{\Theta(\gamma - \gamma_c) - \Theta\left[\gamma - \gamma_m^{\max}(\tau)\right]\Big\}.
\end{eqnarray}
This is the central result of the theory, the general solution of Eq.~\eqref{PDENtildedimless} for spherical grains in $d$ dimensions using random nucleation and growth related rates that contain the fraction of material available for further crystallization of the Kolmogorov-Avrami-Mehl-Johson model [Eqs.~\eqref{ratesdimless}]. The function $\sigma_m(\gamma,\tau)$ appearing in Eq.~\eqref{GenSol1} is a solution of Eq.~\eqref{eqsigmami} and is given by Eqs.~\eqref{sigmam}. The maximal grain size $\gamma_m^{\max}$ is obtained in Eq.~\eqref{gammax} below.

This result has several general properties. First, the solutions can be divided into classes characterized by the dimensionality of the crystallization process $d$, and the time-decay of the nucleation and growth rates, which are specified by the values of $n$ and $m$. Thus, each triplet of non-negative integers $(d,n,m)$ defines another class of the general solution. In the RNG process discussed in the present paper, $n$ is identified with the dimension $d$ and, therefore, the classes are defined by the doublet $(n=d,m)$. Second, the time dependence of Eq.~\eqref{GenSol1} appears through $\sigma_m(\tau)$ [Eq.~\eqref{sigmagen}] and $\gamma_m^{\max}(\tau)$ [Eq.~\eqref{gammax}], which are non-trivial functions of $\tau$. Third, the nucleation rate $I_0$ appears explicitly only in the prefactor $C_d$, whereas the growth rate $v_0$ is present in $\sigma_m$ and  $\gamma_m^{\max}$ as well. This is also the case for $m=-1$ (constant growth rate) and agrees with Ref.~\onlinecite{sekimoto}. On the other hand, both critical times $t_{cv}$ and $t_{cI}$ appear in the exponential through the ratio $t_r$. Fourth, the derivation of the grain size distribution provides cutoffs at the radii of the nucleus and the largest grain found in the sample at time $t$. This will be discussed further in the next section. Finally, we emphasize that the explicit analytical solutions were derived here for specific source and growth terms $\mathbf{v}(\mathbf{r},t)$ and ${\cal D}(\mathbf{r},t)$, namely, Eqs.~\eqref{rates}, which are consistent with the KAMJ model. This is important to remember for the discussion of the next sections, especially the result obtained in the limit $t\to \infty$.

Explicit expressions of the GSD can be written for specific classes of solutions $(d,n,m)$, in particular, for the cases of interest for the description of experimental data, namely, $m=0,1$ and $n=d=1,2,3$. For $m=0$ but arbitrary $(d,n)$, Eq.~\eqref{GenSol1} reads as
\begin{subequations}\label{Nm=0n}
\begin{eqnarray}\label{Nm=0na}
N(\gamma,\tau) &=& \frac{C_d}{\gamma^{d-1}}\,
\frac{\exp\left\{(-1)^n\left[t_r^2\ln\alpha_0\right]^{n+1}\right\}}{\alpha_0}\, \times
\left\{\Theta\left(\gamma-\gamma_c\right) - \Theta\left[\gamma - \gamma_0^{\max}(\tau)\right]\right\},
\end{eqnarray}
with
\begin{eqnarray}\label{Nm=0nb}
\alpha_0(\gamma,\tau) &=& \frac{\gamma-\gamma_c}{{\cal V}_0} + e^{-(\tau-\tau_0)/t_r},
\end{eqnarray}
\end{subequations}
and $\gamma_0^{\max}$ is derived below [Eq.~\eqref{gammaxm0}].

The other case of interest is $m=1$ and arbitrary $(d,n)$
\begin{subequations}\label{Nm=1n}
\begin{eqnarray}\label{Nm=1na}
N(\gamma,\tau) = \frac{C_d}{\gamma^{d-1}}  \exp\left[\left(\mbox{erf}^{-1}\alpha_1\right)^2-\left(t_r^2\,\mbox{erf}^{-1}\alpha_1\right)^{n+1}\right] \times \left\{\Theta\left(\gamma-\gamma_c\right) - \Theta\left[\gamma - \gamma_1^{\max}(\tau)\right]\right\},
\end{eqnarray}
with
\begin{eqnarray}\label{Nm=1nb}
\alpha_1  = {\rm erf}\left(\frac{\tau-\tau_0}{t_r}\right) - \frac{2}{\sqrt{\pi}}\,\frac{\gamma-\gamma_c}{{\cal V}_0},
\end{eqnarray}
\end{subequations}
and $\gamma_1^{\max}(\tau)$ is given by Eq.~\eqref{gammaxm1}.

From the above time-dependent expression we obtain the GSD at full crystallization. For $t\to \infty$, Eqs.~\eqref{Nm=0n} obtained for $m=0$ becomes
\begin{eqnarray}\label{Nm=0ntinfty}
N(\gamma,\tau\to\infty) &=&\frac{C_d}{\gamma^{d-1}}\left(\frac{{\cal V}_0}{\gamma-\gamma_c}\right)\,
\exp\left\{(-1)^n\left[t_r^2\ln\left(\frac{\gamma-\gamma_c}{{\cal V}_0}\right)\right]^{n+1}\right\}\nonumber\\
&&\times\Big[\Theta\left(\gamma-\gamma_c\right) - \Theta\left(\gamma - 1\right)\Big].
\end{eqnarray}
The decay of the growth rate is exponential in this case.
Of particular interest is the case is when $n=d=m+1=1$. The distribution then takes the form\cite{bergmann08}
\begin{eqnarray}\label{Nm=0n=1tinfty}
N(\gamma,\tau\to\infty) &=& \frac{1}{2} \left(\frac{I_0}{v_0}\right)\, \left(\frac{{\cal V}_0}{\gamma-\gamma_c}\right)\,
\exp\left\{-\left[t_r^2\ln\left(\frac{\gamma-\gamma_c}{{\cal V}_0}\right)\right]^2\right\} \nonumber\\
&&\times \Big[\Theta\left(\gamma-\gamma_c\right) - \Theta\left(\gamma - 1\right)\Big].
\end{eqnarray}
This result is closely related to the lognormal distribution. As shown in Ref.~\onlinecite{bergmann08} this is seen more clearly by noting that, in general, $\rho_c\ll \rho_0^\infty$, which implies that for $0<\gamma_c\ll\gamma\leq 1$, the equation simplifies to
\begin{eqnarray}\label{Nm=0ntinftyrc}
N_{>} (\gamma,\tau\to\infty)
= \frac{t_{cv} I_0 }{2\rho_0^\infty} \frac{\exp\left\{- \left[t_r^2\,\ln\gamma\right]^2\right\}}{\gamma}\,
\, \Theta\left(1 - \gamma\right).
\end{eqnarray}
The close relation of this result to the lognormal distribution [see Eq.~\eqref{Nlognormal} in Appendix \ref{s:model} for the expression written in physical units] and its application to full crystallization of amorphous silicon were presented in Ref.~\onlinecite{bergmann08}. We emphasize that obtaining a lognormal-like distribution as a solution of a partial differential equation is a quite remarkable result, and we elaborate on its significance in Sec.~\ref{ss:remarks}.

Contrary to the case $m=0$, no significant simplification of Eqs.~\eqref{Nm=1n} ($m=1$) is obtained in the limit $t\to\infty$.
The above expressions allow studying the behavior of the theoretical distribution in the following sections, and were used to analyze experimental data during,\cite{billMRS09} and at full crystallization\cite{bergmann08} of amorphous silicon. 

\subsection{Maximal grain size}\label{ss:rmax}

The solution presented in the previous paragraph contains the maximal grain size $\gamma_m^{\max}(\tau)$ that can be observed in a sample undergoing RNG crystallization. This quantity is obtained from $d\rho = v_m(t)dt$. Integrating one immediately obtains in dimensionless variables
\begin{eqnarray}\label{gammaxeq}
 \gamma_m^{\max}(\tau) \equiv \frac{\rho_m^{\max}(t)}{\rho_m^\infty} &=& \gamma_c + \frac{1}{t_r}\int_{\tau_0}^\tau v_m(\tau')d\tau' = \gamma_c + u_m(\tau_0,\tau).
\end{eqnarray}
For $v_m(\tau)$ given by Eq.~\eqref{vdimless}, the above equation can be written in terms of the upper incomplete gamma functions, $\Gamma[a,x]$,
\begin{eqnarray}\label{gammax}
\gamma_m^{\max}(\tau) &=& 1 - \frac{{\cal V}_0}{m+1}\,\Gamma\left[\frac{1}{m+1},\left(\frac{\tau-\tau_0}{t_r}\right)^{m+1} \right].
\end{eqnarray}
Explicit expressions of Eq.~\eqref{gammax} with $m=0$ and $m=1$ are
\begin{subequations}\label{gammaxm}
\begin{eqnarray}\label{gammaxm0}
\gamma_0^{\max}(\tau) &=& 1 - {\cal V}_0 e^{-(\tau-\tau_0)/t_r}, \quad m=0,\\
\label{gammaxm1}
\gamma_1^{\max}(\tau) &=& 1 - {\cal V}_0\, \frac{\sqrt{\pi}}{2}\,{\rm erfc}\left(\frac{\tau-\tau_0}{t_r}\right), \quad m=1,
\end{eqnarray}
\end{subequations}
with ${\rm erfc} = 1-{\rm erf}$.

It is instructive to digress from the main path of the paper and determine the maximal grain size in the case $m=-1$, which corresponds to a constant growth rate $v_{-1}(t) = e\,v_0$. We obtain
\begin{eqnarray}
\gamma_{-1}^{\max} (\tau) = \gamma_c + e {\cal V}_0 \left(\frac{\tau-\tau_0}{t_r}\right).
\end{eqnarray}
We note that $\gamma_{-1}^{\max} (\tau\to\infty)\to\infty$. This is consistent with previous work that showed unbounded maximal grain size but is not physically justified unless coalescence is taken into account. We do not analyze this case further here, as we would have to redefine our dimensionless grain size $\gamma$.

One can also determine the maximal grain size $\rho_m^\infty$ once crystallization is completed ($t\gg \max\{t_{cv}, t_{cI}\}$). In the dimensionless formalism used in this section this corresponds to $\gamma_m^\infty=1$. The expression for $\rho_m^\infty$ can be transformed into an interesting relation between fundamental parameters of the model for any $m$ [see Eq.~\eqref{ginf} in Appendix \ref{a:solutioninrandt}]
\begin{eqnarray}\label{paramrelation}
\frac{\rho_m^\infty - \rho_c}{t_{cv}v_0} = \Gamma\left[\frac{m+2}{m+1}\right].
\end{eqnarray}
This equation can be used as a self-consistency check or to determine the value of one parameter once the others have been measured.\\

\begin{figure}[h]
\includegraphics[width=0.5\textwidth,clip=true]{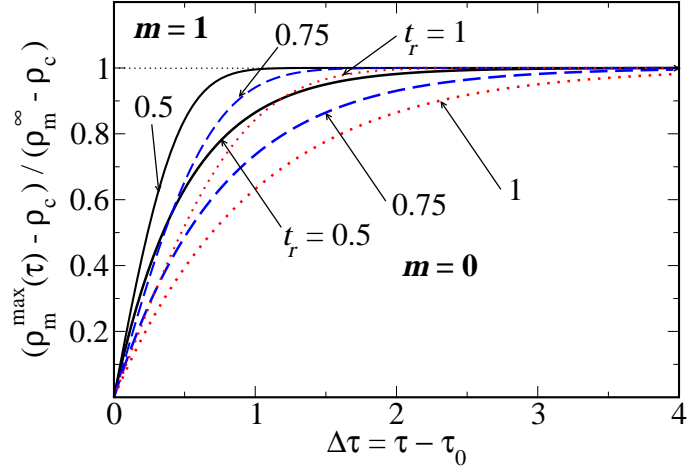}
\unitlength1cm
\caption{\label{fig:gammatau}Maximal grain size as a function of time $(\gamma_m^{\max}(\tau) - \gamma_c)/(1 - \gamma_c) =(\rho_m^{\max}(\tau)- \rho_c)/(\rho_m^{\infty}- \rho_c)$. The normalization is such that the curves do not depend on $\rho_m^\infty {\cal V}_0 = t_{cv}v_0$ (see text).  The thick lines are for $m=0$ and the thin lines $m=1$. For each set, we have $t_r = 0.5$ (black solid), $t_r = 0.75$ (blue dashed), and $t_r=1$ (red dotted). The saturation of $\gamma_m^{\max}(\tau)$ is shown to occur at earlier times with decreasing $t_r$ and/or increasing $m$.}
\end{figure}
The time-dependence of the maximal grain size is depicted in Fig.~\ref{fig:gammatau} for $m=0,1$ and reasonable values of $t_r$.\cite{billMRS09} Within our model $v_m(\tau)$ given by Eq.~\eqref{vdimless} is proportional to ${\cal V}_0$. Hence, if follows from Eqs.~\eqref{gammax}, \eqref{gammaxrt}, and \eqref{ginf} that $\left(\gamma_m^{\max}(\tau) - \gamma_c\right)/(1 - \gamma_c) = \left(\rho_m^{\max}(\tau) - \rho_c\right)/(\rho_m^\infty - \rho_c)$ does {\it not} depend on ${\cal V}_0$; the results in Fig.~\ref{fig:gammatau} depend only on $m$ and $t_r$. It is also worth pointing out that, except for early stages of crystallization, the inequality $\rho_m^{\max}(\tau) \gg \rho_c$ is generally satisfied and Fig.~\ref{fig:gammatau} essentially displays the ratio $\rho_m^{\max}(\tau)/\rho_m^\infty$. The figure demonstrates that the radius of the largest grain observed in the sample saturates rapidly in time to reach the size observed at full crystallization (calculations of the GSD for the parameters of Fig.~\ref{fig:gammatau} show that full crystallization is obtained for $\Delta\tau\gtrsim 8$ if $m=0$ and $\Delta\tau \gtrsim 4$ for $m=1$.) A decreasing value of $t_r$ ({\it e.g.}, faster decay of the growth rate at constant nucleation rate) enhances the rate at which $\rho_m^{\max}$ reaches $\rho_m^\infty$. A similar conclusion is reached with increasing $m$. This behavior results from the fact that both a decrease of $t_{cv}$ and increase of $m$ accentuate the time decay of the growth rate, which --as discussed in previous sections-- reflects the impingement caused by neighboring grains.

The analytical determination of the maximal grain size in terms of fundamental parameters of the model is of particular interest for those interested in a probabilistic approach using computer simulation to determine the grain size distribution because, as discussed in Ref.~\onlinecite{redner90}, the proper account of rare events (the size of the largest and smallest grains) is essential for describing the data with the adequate distribution and obtaining the correct value for its average.\\

\section{Characterization and time-evolution of $N(\gamma,\tau)$}\label{s:timeevolution}

This section is devoted to the characterization of the grain size distribution obtained in Eq.~\eqref{GenSol1}. First we analyze the influence of the model parameters $t_r$ and $v_0$, the ratio of critical times and the growth rate, respectively. This leads to scaling properties for $v_0$ in the limit $\tau\to\infty$. We then consider the time-evolution of the distribution. Finally, we show how for certain classes of solutions $(n,m)$ the distribution takes the lognormal form in the asymptotic limit of large time.

To proceed with the numerical analysis, it is appropriate to consider the normalized GSD
\begin{subequations}\label{Nnormalized}
\begin{eqnarray}
\bar{N}(\gamma,\tau) = \frac{N(\gamma,\tau)}{N(\tau)},
\end{eqnarray}
where the total number of grains at time $\tau$ is given by
\begin{eqnarray}
N(\tau) = \int_0^\infty\,N(\gamma,\tau) \,d\gamma = \int_{\gamma_c}^{\gamma_m^{\max}(\tau)} N(\gamma,\tau)\,d\gamma.
\end{eqnarray}
\end{subequations}
This normalization procedure eliminates the constant factor $C_d$ defined in Eq.~\eqref{Cd}. As mentioned earlier, this is the only term containing the nucleation rate $I_0$ and the results discussed in the remainder of this paper are therefore independent of the explicit value of $I_0$.  Such normalization is also useful for comparing the theory with experimental data.\cite{bergmann08,billMRS09}

The figures in the present paper all depict the normalized GSD $\bar{N}(\gamma,\tau)$. Thus, the area under the curve is one. It is also worth noting for the next sections that none of the distributions discussed here contain divergences.

\subsection{Moments and maxima of the distribution}\label{ss:moments}

The time-dependence of the distribution can be characterized in terms of its moments. In particular, the first three moments, which give the mean $\mu$, the variance $\sigma^2$, and the skewness $\gamma_1$ of the distribution. $\sigma^2$, $\gamma_{1}$ are central moments and the latter is normalized. The two higher moments give an indication about the spread and asymmetry of the distribution about the mean.  All moments are calculated for the normalized GSD [Eq.~\eqref{Nnormalized}] and thereby independent of the nucleation rate coefficient $I_0$. We define
\begin{subequations}\label{moments}
\begin{eqnarray}\label{mom1}
\mu(\tau) &=& \int_0^\infty d\gamma\, \gamma\,\bar{N}(\gamma,\tau),\\
\label{mom2}
\sigma^2(\tau) = \mu_{2}(\tau) &\equiv& \int_0^\infty d\gamma\, \left[\gamma-\mu(\tau)\right]^2 \bar{N}(\gamma,\tau),\\
\label{mom3}
\gamma_{1}(\tau) \equiv \frac{\mu_{3}}{\mu_{2}^{3/2}} &=& \frac{1}{\mu_{2}^{3/2}} \int_0^\infty d\gamma\, \left[\gamma-\mu(\tau)\right]^3 \bar{N}(\gamma,\tau).
\end{eqnarray}
\end{subequations}
The definitions are given for the dimensionless GSD written in terms of $\gamma$. Moments for $N(\rho,t)$ are given in Appendix \ref{a:solutioninrandt}. The conventional notation for the third moment, $\gamma_{1}$ (always written with its index), should not be confused with the variable $\gamma = \rho/\rho_m^\infty$ (never written with an index.) We calculate the time dependence of the mean and variance for the case $m=0$ and $n=1,2,3$ in Sec.~\ref{ss:tevol}.

In some cases, it may be of interest to compare the radius $\gamma_{\max,i}^{n,m}$ ($i=1,2,\dots$) for which the distribution is maximal to the mean $\mu$ of the GSD. Since it turns out that under certain circumstances the GSD [Eq.~\eqref{GenSol1}] has more than one maximum (see Sec.~\ref{ss:tevol}), we add the index $i$ to $\gamma_{\max}^{n,m}$. We thus define $\gamma_{\max,i}^{n,m}$ by
\begin{eqnarray}
\left. \frac{\partial N(\gamma,\tau)}{\partial \gamma}\right|_{\gamma_{\max,i}^{n,m}} = 0,\quad \left.\frac{\partial^2 N(\gamma,\tau)}{\partial \gamma^2}\right|_{\gamma_{\max,i}^{n,m}} < 0.
\end{eqnarray}
The analytical form of $\gamma_{\max,m}^{n,m}$ can be determined from the zero of the derivative
\begin{eqnarray}
\left(\frac{\sigma_m - \tau_0}{t_r}\right)^{m+1} - t_r^{n+1} \left(\sigma_m - \tau_0\right)^{n+1} = (d-1)\gamma.
\end{eqnarray}
For the case $n=d=1$ and $m=0$ the maximum in the open interval $\left(\gamma_c,\gamma_m^{\max}(\tau)\right)$ is given by
\begin{eqnarray}
\label{peakposm=0}
\gamma_{\max}^{n=1,m=0}(\tau) = \gamma_c + {\cal V}_0\left[e^{-1/2t_r^4} + e^{-(\tau-\tau_0)/t_r}\right].
\end{eqnarray}

\subsection{Grain size distribution at $t\to\infty$}\label{ss:tinfty}

\begin{figure}[h]
\includegraphics[width=0.45\textwidth,clip=true]{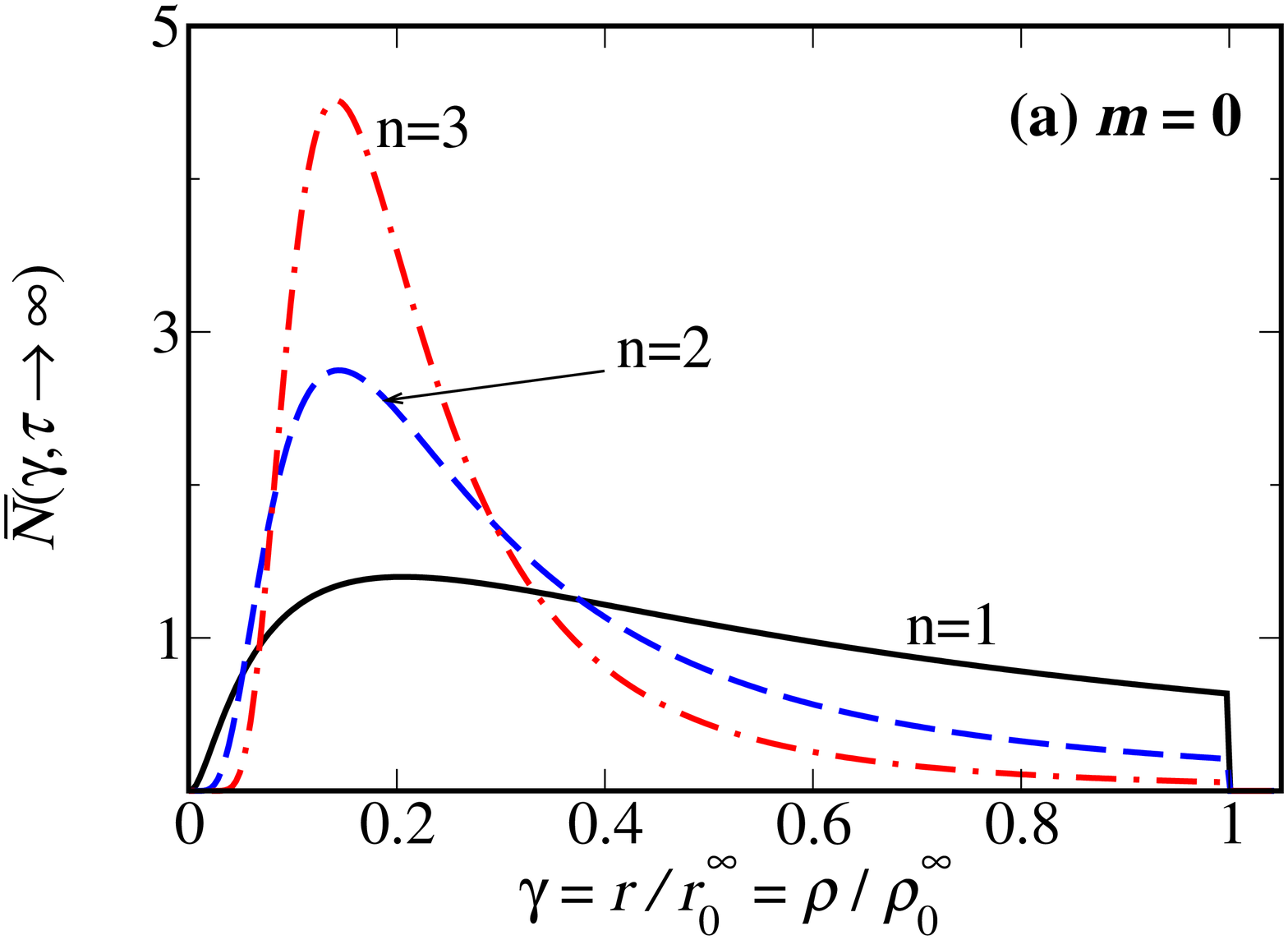}
\hspace*{.8cm}
\includegraphics[width=0.45\textwidth,clip=true]{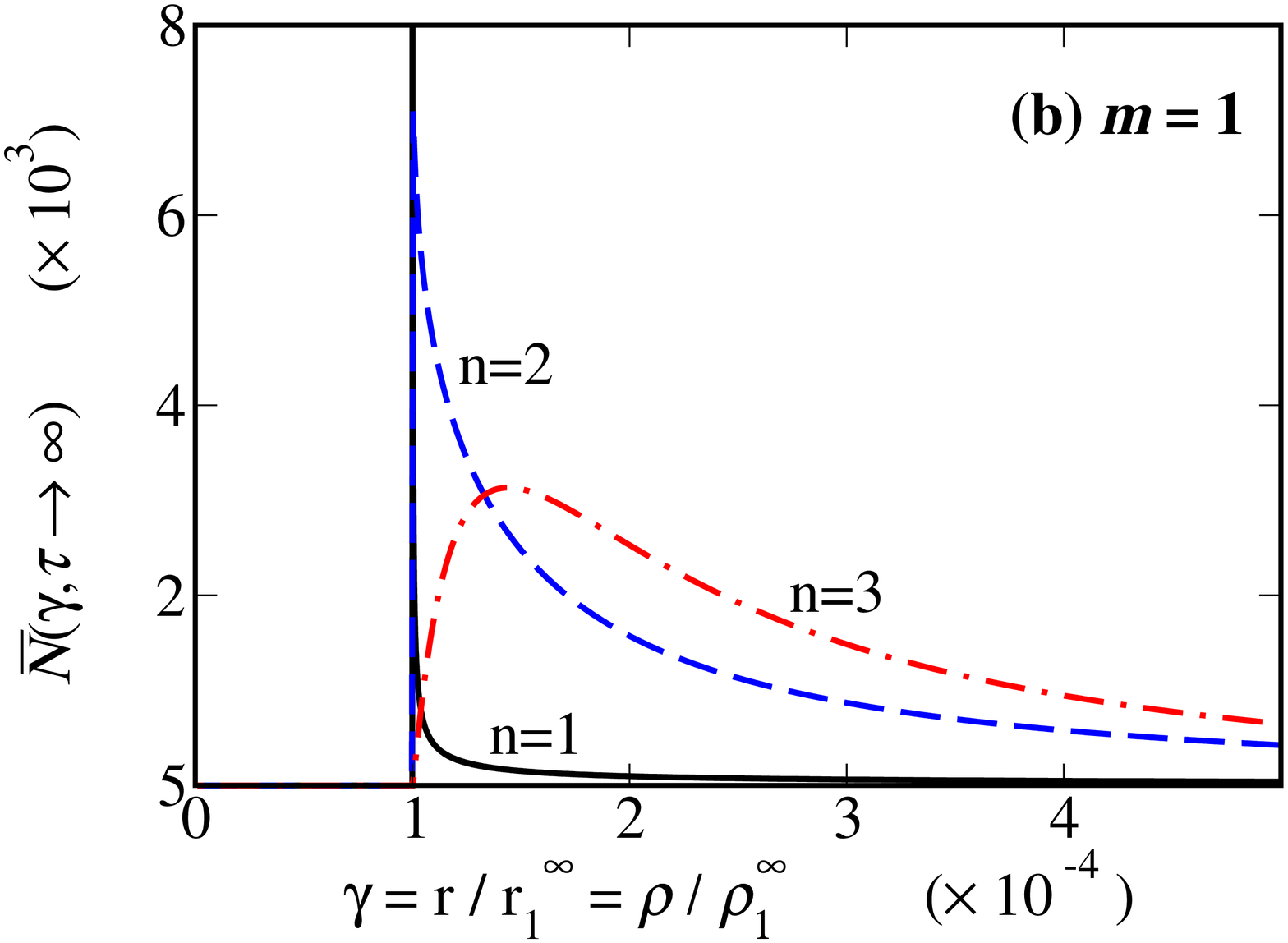}
\begin{center}
\unitlength1cm
\begin{picture}(0,0)
\end{picture}
\caption{\label{fig:Nm0n}Normalized distribution $\bar{N}(\gamma,\tau\to\infty)$ for $n=1$ (black solid), $n=2$ (blue dashed), $n=3$ (red dash-dotted line). (a) $m=0$. (b) $m=1$. The abscissa is the normalized grain radius $\gamma = \rho/\rho_m^\infty$, which is independent of $n$  (see text). In this and the following figures we consider generic parameters chosen in the range of interest to the experiment:\cite{billMRS09} $t_r = 0.75$, ${\cal V}_0=1$.}
\end{center}
\end{figure}

For $t\to\infty$, the (unormalized) distribution is given by Eq.~\eqref{Nm=0ntinfty} for $m = 0$. We remind that in this case the time-decay of the growth rate is exponential. The normalized distribution $\bar{N}(\gamma,\tau)$ is shown in Fig.~\ref{fig:Nm0n}(a) for $n=d=1,2,3$ and the parameter values specified in the caption. These are physically reasonable choices of parameters as discussed in Refs.~\onlinecite{bergmann08} and \onlinecite{billMRS09}.
All curves have identical upper cutoff (the upper cutoff is out of the range of the figure for $m=1$; Fig.~\ref{fig:Nm0n} right). For $m=0$ both the mean value and the variance of the distribution decrease with larger $n$;
the peak of the distribution is sharper and shifts to lower values of $\gamma$ for increasing $n$. Since we identify $n$ with the dimensionality of the system, we expect the GSD of three-dimensional crystallization to be sharper than that of thin films when the thickness of the film is smaller than the average grain size at full crystallization. Furthermore, the majority of grains have smaller size in three dimensions than in two dimensions.
 
The distribution is also affected by the choice of $m$. Choosing $m=1$ (Fig.~\ref{fig:gammatau}, right) instead of $m=0$ (Fig.~\ref{fig:gammatau}, left) implies a faster time-decay of the growth rate.
From Fig.~\ref{fig:gammatau} we obtain the physically intuitive result that a stronger time-decay of the growth rate leads to sharper peaks, with a maximum located at smaller grain radii. This is emphasized by the difference in abscissa and ordinate scales between  $m=0$ and $m=1$ in Fig.~\ref{fig:Nm0n}. Replacing the exponential decay ($m=0$) of the growth rate by a Gaussian decay ($m=1$) has a dramatic effect on the grain size distribution. This results in the unambiguous choice $m=0$ to describe the experimental data of solid-phase crystallization of amorphous silicon.\cite{bergmann08,billMRS09}

\subsection{Dependence of $N(\gamma,\tau)$ on $t_r$ and ${\cal V}_0$}\label{ss:trv0}

The grain size distribution $\bar{N}(\gamma,\tau)$ is a function of $d, n, m, t_r, t_0$ and $v_0$. However, not all parameters need always be known to determine the GSD.
Considering Eqs.~\eqref{sigmagen}, \eqref{gammax}, and \eqref{relationV0gamc}, we note that it is possible to reduce the number of parameters that appear in Eq.~\eqref{GenSol1}. For example, from Eqs.~\eqref{sigmagen} and \eqref{gammax}, we can express $\sigma_m(\gamma,\tau)$ in terms of $t_r$, $\gamma_c$, $\gamma_m^{\max}(\tau)$, and $\tau_0$. Thus, the normalized GSD can be written in terms of the latter parameters, and ${\cal V}_0$ and ${\cal I}_0$ are not explicitly needed in the expression. Conversely, it is possible to express the GSD in terms of the two latter quantities, thereby, removing other parameters. For example, for $m=0$, Eq.~\eqref{relationV0gamc} implies ${\cal V}_0 = (1-\gamma_c)\Gamma^{-1}\left[(m+2)/(m+1)\right]$, which can be used to write $(\gamma-\gamma_c)/{\cal V}_0 = \Gamma\left[(m+2)/(m+1)\right] (\gamma-\gamma_c)/(1-\gamma_c)$ in Eqs.~\eqref{Nm=0n} and \eqref{Nm=0ntinfty}. As is often the case, $\gamma_c \ll 1$, which implies that the ratio is essentially $\gamma$. Then, in the limit $\tau\to\infty$, the GSD [Eq.~\eqref{Nm=0ntinfty}] only depends on $t_r$ and $C_d$ and the normalized GSD only on $t_r$. At finite times $\gamma_m^{\max}(\tau)$ remains present and is directly proportional to ${\cal V}_0$. However, $\gamma_m^{\max}$ can be determined experimentally, and ${\cal V}_0$ is not required. To summarize, the choice of which parameters are needed and which can be obtained from the expressions above depends on the particular situation under consideration. A natural choice of parameters to discuss the properties of the normalized distribution is $t_r$ and ${\cal V}_0$. We discuss in this section the general dependence of $\bar{N}(\gamma,\tau)$ on these two parameters.

\begin{figure}[h]
\includegraphics[width=0.35\textwidth]{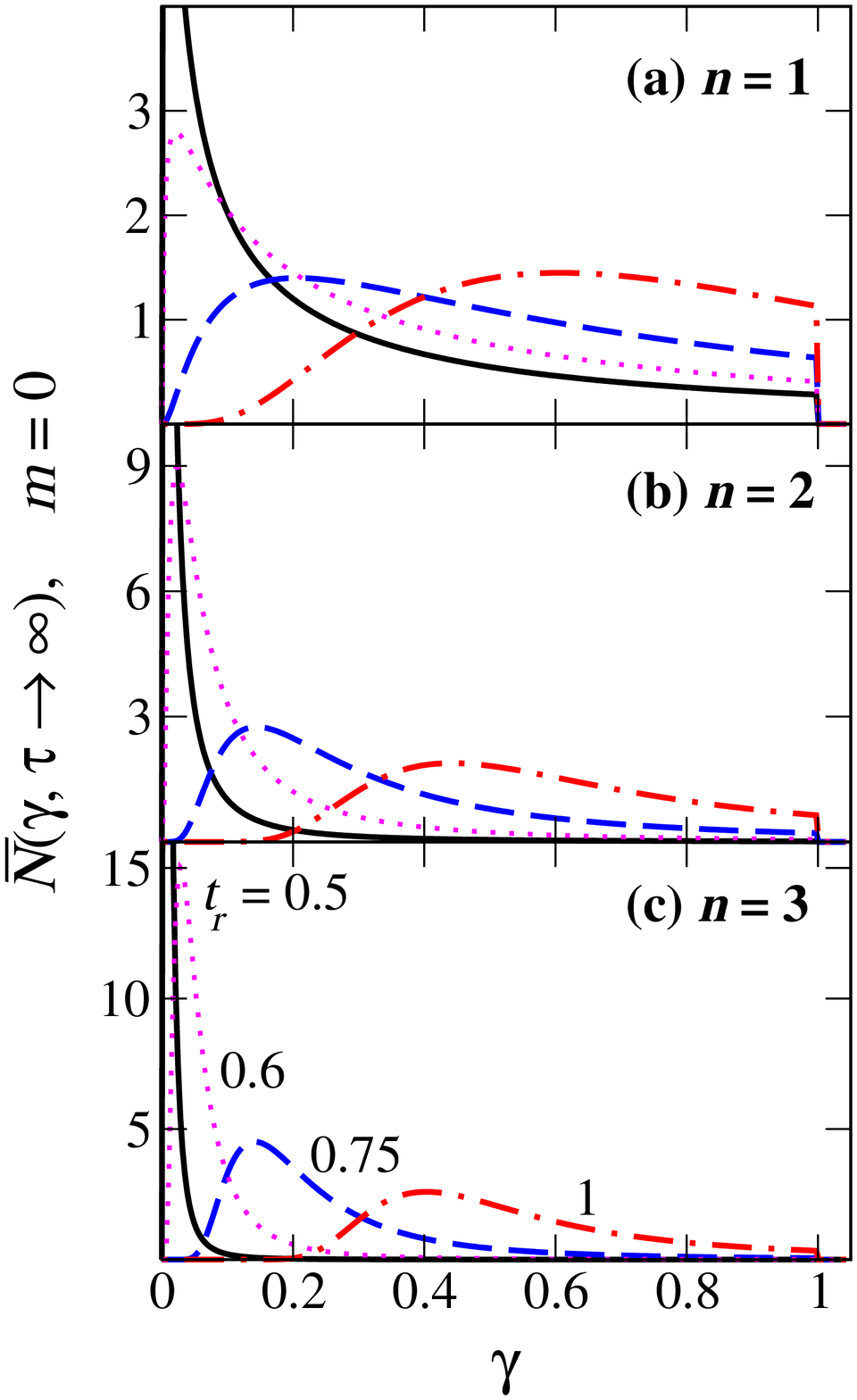}
\hspace*{0.2cm}
\includegraphics[width=0.367\textwidth]{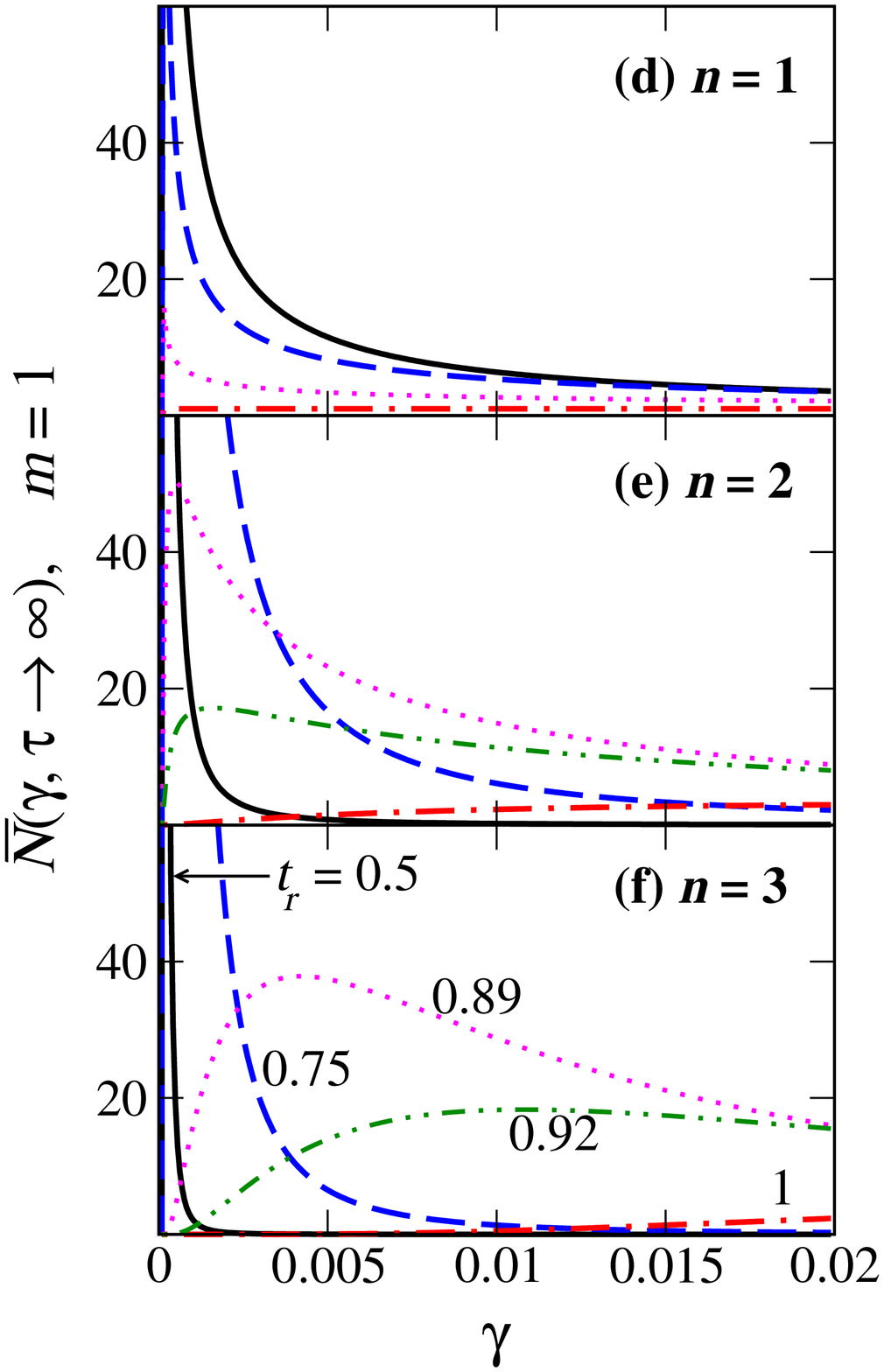}
\begin{center}
\caption{\label{fig:m0nVtr}$\bar{N}(\gamma,\tau\to\infty)$ for different critical time ratios $t_r=\sqrt{t_{cv}/t_{cI}} = 0.5$ (black solid line), $t_r = 0.75$ (dashed blue line), and $t_r = 1$ (dash-dotted red line). In Figs.~(a,b,c) the magenta dotted line is for $t_r = 0.6$. In Figs.~(d,e,f) the magenta dotted line is for $t_r = 0.89$ and the green dash-double dotted line is for $t_r = 0.92$. The left column is for $m=0$ and the right column for $m=1$. Rows are for $n=1$ (Figs.~a,d), $n=2$ (b,e), and $n=3$ (c,f). Note the different scales on the abscissa and the ordinates between the two columns. For all figures ${\cal V}_0 = 1$.}
\end{center}
\end{figure}

\begin{figure}[h]
\includegraphics[width=0.35\textwidth]{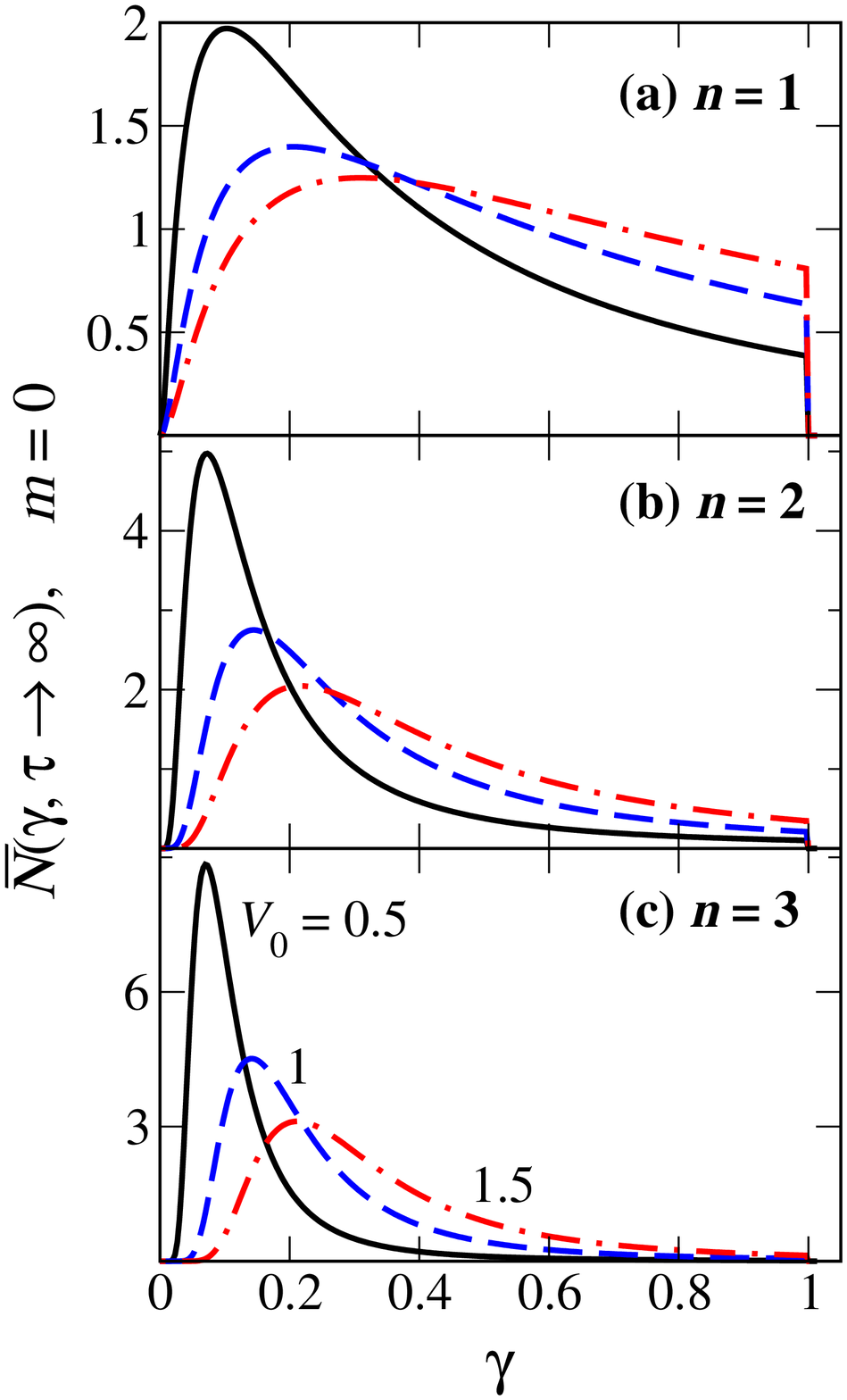}
\hspace*{0.2cm}
\includegraphics[width=0.367\textwidth]{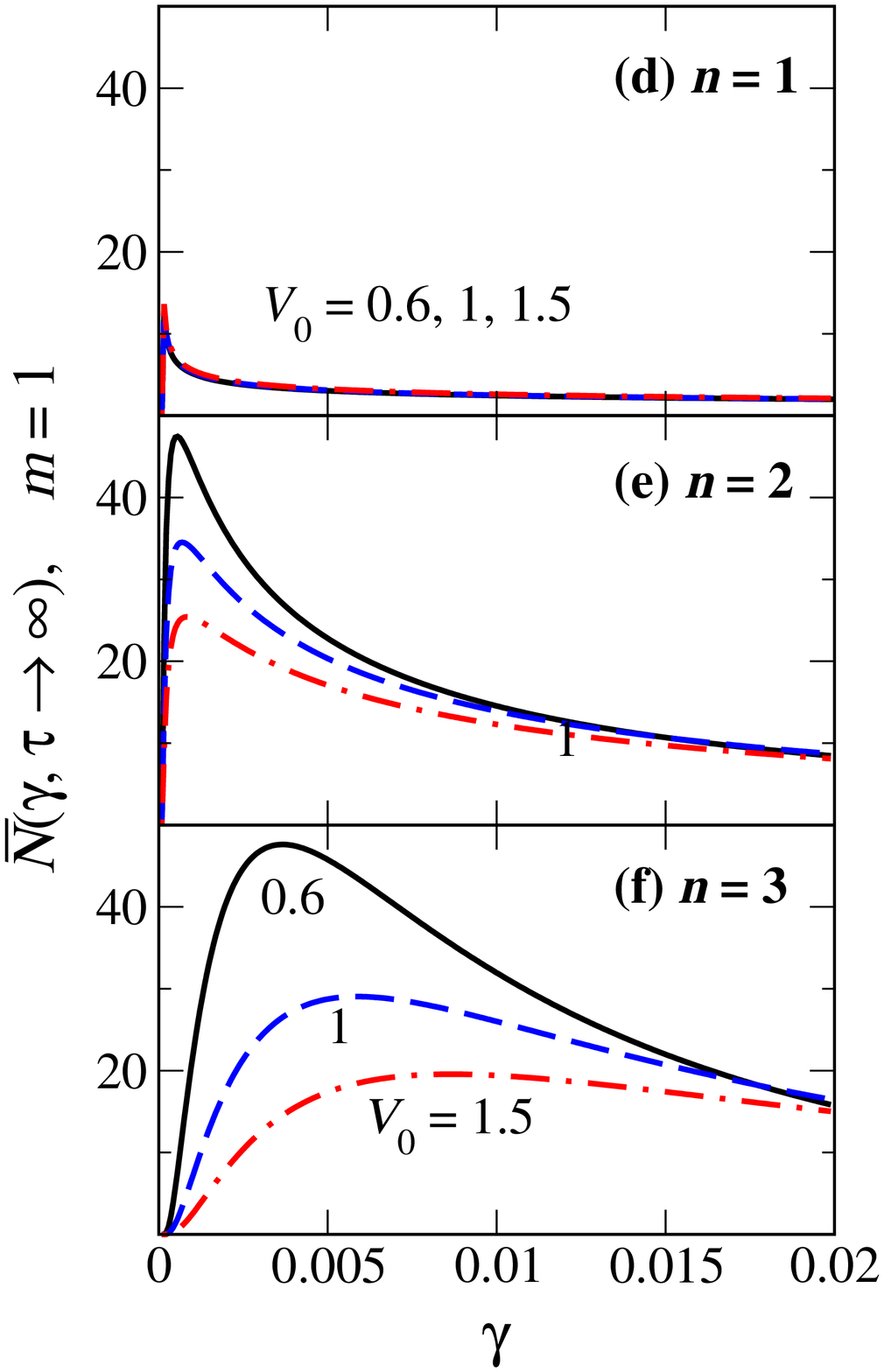}
\begin{center}
\caption{\label{fig:m0nVv0} $\bar{N}(\gamma,\tau\to\infty)$ for different growth rate amplitudes ${\cal V}_0$. Black solid line: ${\cal V}_0 = 0.5$ (left column) and 0.6 (right column). Dashed blue line: ${\cal V}_0 = 1$. Dash-dotted red line: ${\cal V}_0 = 1.5$. Left column: $m=0$. Right column: $m=1$. Figs.~(a,d) are for $n=1$, (b,e) for $n=2$, and (c,f) for $n=3$. In Figs.~(a,b,c) $t_r = 0.75$, and in Figs.~(d,e,f) $t_r = 0.9$.}
\end{center}
\end{figure}

To study the generic influence of $t_r$ on the behavior of $\bar{N}(\gamma,\tau)$, it turns out to be sufficient to consider Eq.~\eqref{GenSol1} in the limit $t\to\infty$ when full crystallization is achieved. This limit has been calculated analytically for classes of solutions $(d,m,n)$ relevant for experimental studies in Eqs.~(\ref{Nm=0n}-\ref{Nm=0ntinfty}). Figures \ref{fig:m0nVtr} and \ref{fig:m0nVv0} highlight the influence of $t_r$ and $v_0$, respectively, on the shape of the distribution for $m=0$ (left column) and $m=1$ [right column; the rows are for $n=1$ (top), $n=2$ and $n=3$ (bottom)].
The first general observation is that in all cases, the distribution displays one maximum, and cutoffs at $\gamma=\rho_c/\rho_m^{\infty}$ and $\gamma = 1$. While the latter is obvious from Eq.~\eqref{GenSol1}, the former is only true at large times within or model, as will be seen in the next section.

As $t_r = \sqrt{t_{cv}/t_{cI}}\lesssim 1$ decreases, the number of small grains increases at the expense of the formation of larger grains. This is confirmed in Fig.~\ref{fig:m0nVtr}, which shows that irrespective of the value of $n$ and $m$ a decrease of $t_r$ results in an increase in amplitude, a sharpening of the peak, and a shift of the maximum to smaller values of $\gamma$. Furthermore, the properties observed on Fig.~\ref{fig:Nm0n} are also found in Fig.~\ref{fig:m0nVtr}. Thus, the qualitative features inferred from Fig.~\ref{fig:Nm0n} do not strongly depend on the particular value of $t_r$ when taken within a physically reasonable range.
It is interesting to observe that the position of the maximum of the distribution for $n=2,3$ and $m=1$ is essentially insensitive to the value of $t_r$ unless the latter is very close to one. Even then, comparing the order of magnitude of abscissa and ordinate scales of Figs.~\ref{fig:m0nVtr}(e) and \ref{fig:m0nVtr}(f) and \ref{fig:m0nVtr}(a) and \ref{fig:m0nVtr}(b), the peak of the distribution barely shifts for $m=1$ and increasing $t_r$. On the other hand, the amplitude of the maximum strongly varies with $t_r$.
Finally, the distribution for $n=m=1$ and $t_r=1$ is rectangular [lower red dotted line in Fig.~\ref{fig:m0nVtr}(d)]. This result is derived in the next section.

We now study the influence of the growth rate amplitude ${\cal V}_0$ on the grain size distribution. We can also limit our considerations to the limit $\tau\to \infty$. The influence of the growth rate amplitude ${\cal V}_0$ is shown in Fig.~\ref{fig:m0nVv0}.  Physically, an increase of ${\cal V}_0$ implies a faster growth of the grains in a given time interval, thus, leading overall to the presence of larger crystallized grains in the sample. According to Fig.~\ref{fig:m0nVtr}, the ratio $t_r$ determines how sharp the peak is at given ${\cal V}_0$. At fixed value of $t_r$ and for all $n$, an increase of ${\cal V}_0$ results in a decrease and broadening of the peak as well as a stretching of the distribution to higher values of $\gamma$.
As in Fig.~\ref{fig:m0nVtr}, the peak for $m=1$ is also barely shifting with change of ${\cal V}_0$, as opposed to the case $m=0$. Although a trend similar to Fig.~\ref{fig:m0nVtr} is observed when increasing $n$, the growth and shift of the peak are less pronounced. Note that the blue dashed curves with parameters $t_r = 0.75$ and ${\cal V}_0 = 1$ are the same in Figs.~\ref{fig:m0nVtr} and \ref{fig:m0nVv0}, which allows a scaled comparison between of the two figures.

To conclude this section, Figs.~\ref{fig:m0nVtr} and \ref{fig:m0nVv0} show that the behavior for $m=1$ as a function of $t_r$ and ${\cal V}_0$ is similar to that for $m=0$, but the peaks are much sharper, larger in magnitude (we remind that none of the distributions contain divergences), and located at smaller values of $\gamma$. In addition, we note that the shift of the peak with increasing value of $t_r$ or ${\cal V}_0$ is much less pronounced for $m=1$ than for $m=0$. This reflects the difference between the exponential-type decay ($m=0$) and the Gaussian decay ($m=1$) of the growth rate. It also reflects the presence of the logarithm or the error function of $\gamma$, respectively, in the distribution. The different behavior of the distribution for $m=0$ and $m=1$ is distinctive enough to determine which distribution is most suited for a set of experimental data. For example, in our analysis of crystallization in amorphous Si thin films (Refs.~\onlinecite{bergmann08} and \onlinecite{billMRS09}), one is unambiguously led to choose $m=0$. Remember finally that in the KAMJ model, $n=d$ is the dimensionality of the crystallization process. Thus, the fit of experimental data for different values of $n$ could be used to determine the dimensionality of the crystallization process.\\
 
\subsection{Time-evolution of the distribution}\label{ss:tevol}

In the previous section, we discussed how model parameters influence the grain size distribution in the fully crystallized limit $t\to\infty$. The results qualitatively hold for all times $\tau_0\leq \tau <\infty$ as well. We now study how the distribution evolves in time for fixed values of the parameters as the RNG process takes place, starting from an amorphous solid in $d$ dimensions.

It is instructive to consider first two simple cases of the general solution. When the critical times for nucleation and growth are identical, $t_{cv} = t_{cI} = t_c$ (implying $t_r = 1$), Eq.~\eqref{GenSol1} reduces to
\begin{eqnarray}\label{GenSoltc}
N(\gamma ,\tau) &=&
\frac{C_d}{\gamma^{d-1}} e^{\left[(\sigma_m -\tau_0)^{m+1} - (\sigma_m-\tau_0)^{n+1}\right]}\times
\Big\{\Theta(\gamma - \gamma_c) - \Theta\left[\gamma - \gamma_m^{\max}(\tau)\right]\Big\}.
\end{eqnarray}
Assuming further that $n=m$, we obtain the simple form
\begin{eqnarray}
\label{GenSoltn=mtr=1}
N(\gamma ,\tau) =
\frac{C_d}{\gamma^{d-1}} \times \Big\{\Theta(\gamma - \gamma_c) - \Theta\left[\gamma - \gamma_m^{\max}(\tau)\right]\Big\}.
\end{eqnarray}
This latter case is depicted in Fig.~\ref{fig:m0nVtr}(d) (red dash-dotted bottom line, close to the abscissa) and approximates for arbitrary values of $n$ and $m$ the grain size distribution at times very close to the incubation time $t_0$.\cite{bergmann08} However, our description of the nucleation process [Eq.~\eqref{Ddelta}] is rather coarse and is not expected to reproduce accurately the early stages of crystallization.\cite{shinucleation}

Comparing the above two expressions, we note that in the first case the distribution displays an exponential dependence of time [remember that $\sigma_m = \sigma_m(\gamma,\tau)$], whereas in the second case the distribution is constant in time. These two examples emphasize the fact that the exact functional form of nucleation and growth rates, and, in particular, their time dependence [the choice of $n$ and $m$ in the power laws of Eq.~\eqref{ratesdimless}], is critical in determining the grain size distribution $N(\rho,t)$ at all times, including in the asymptotic regime $t\to \infty$.

To analyze the time dependence, we focus on the classes of solution defined by $m=0,1$ and $n=1,2,3$, where analytical solutions of Eq.~\eqref{PDEgeneral1} can be derived, and that also appear to be most relevant for the description of experimental data.\cite{bergmann08,billMRS09} The  expressions for the grain size distribution are given by Eqs.~\eqref{Nm=0n} for $m=0$ and Eq.~\eqref{Nm=1n} for $m=1$.

\begin{figure}[h]
\includegraphics[width=0.4\textwidth]{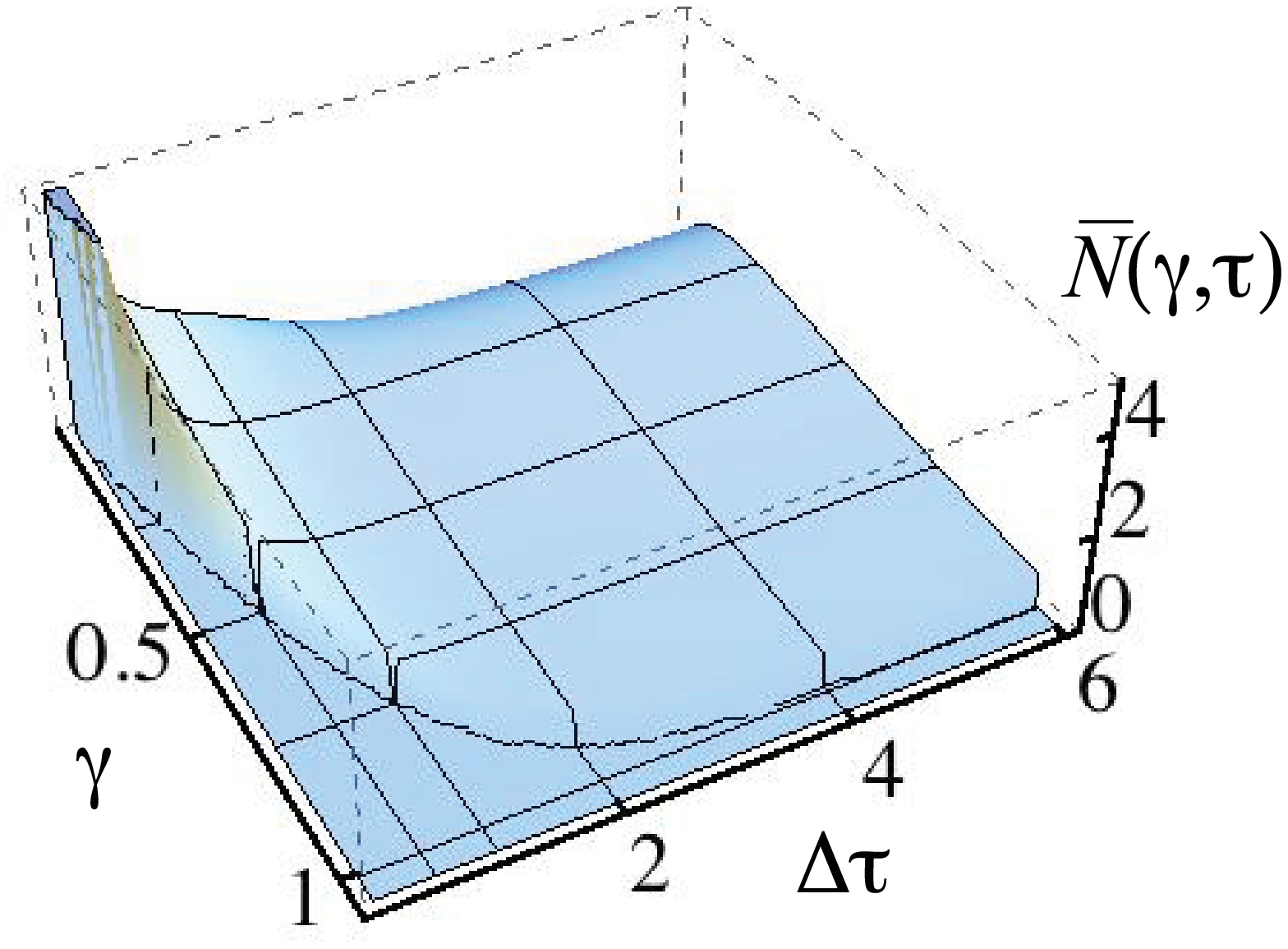}
\includegraphics[width=0.3\textwidth]{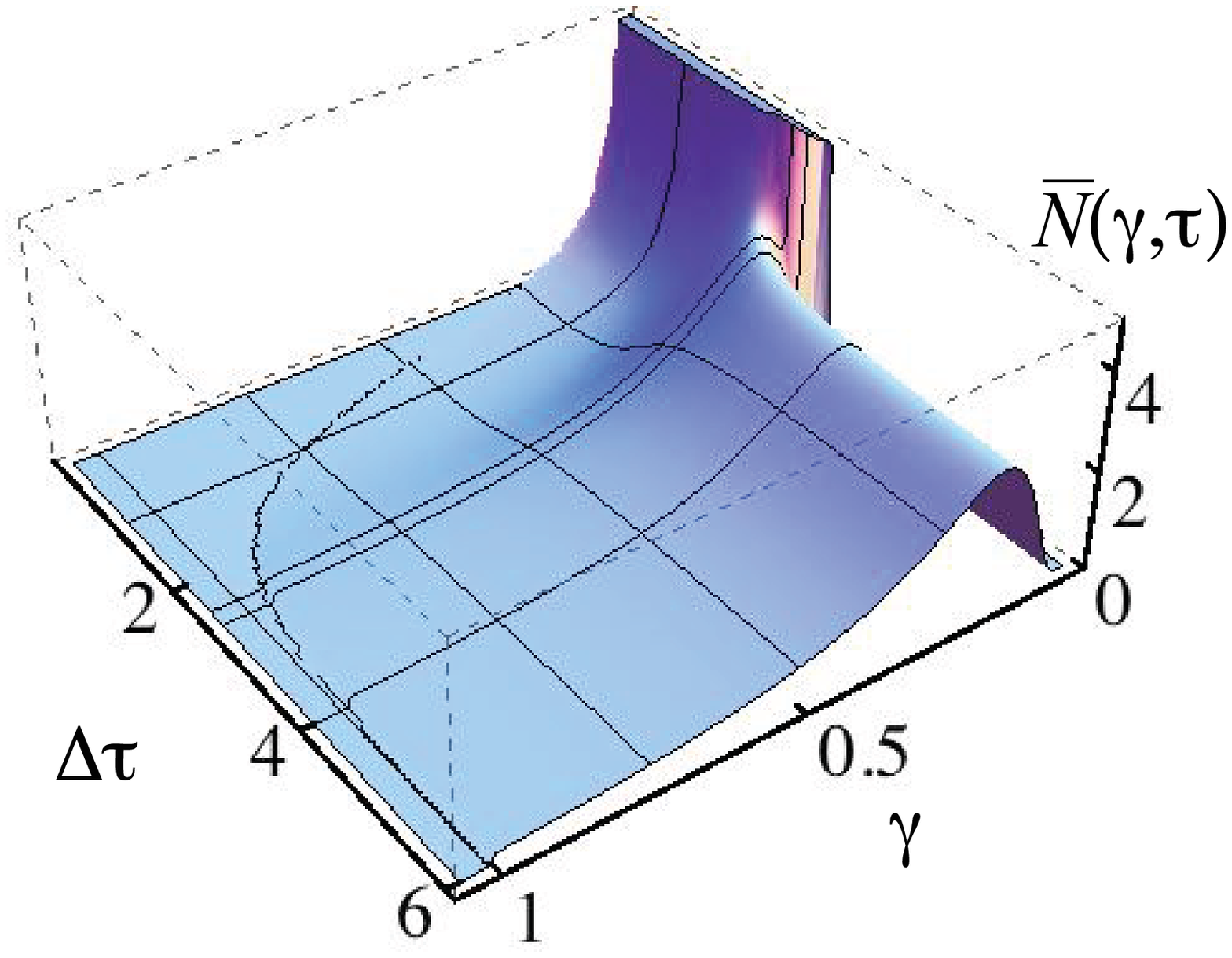}\\
\includegraphics[width=0.35\textwidth]{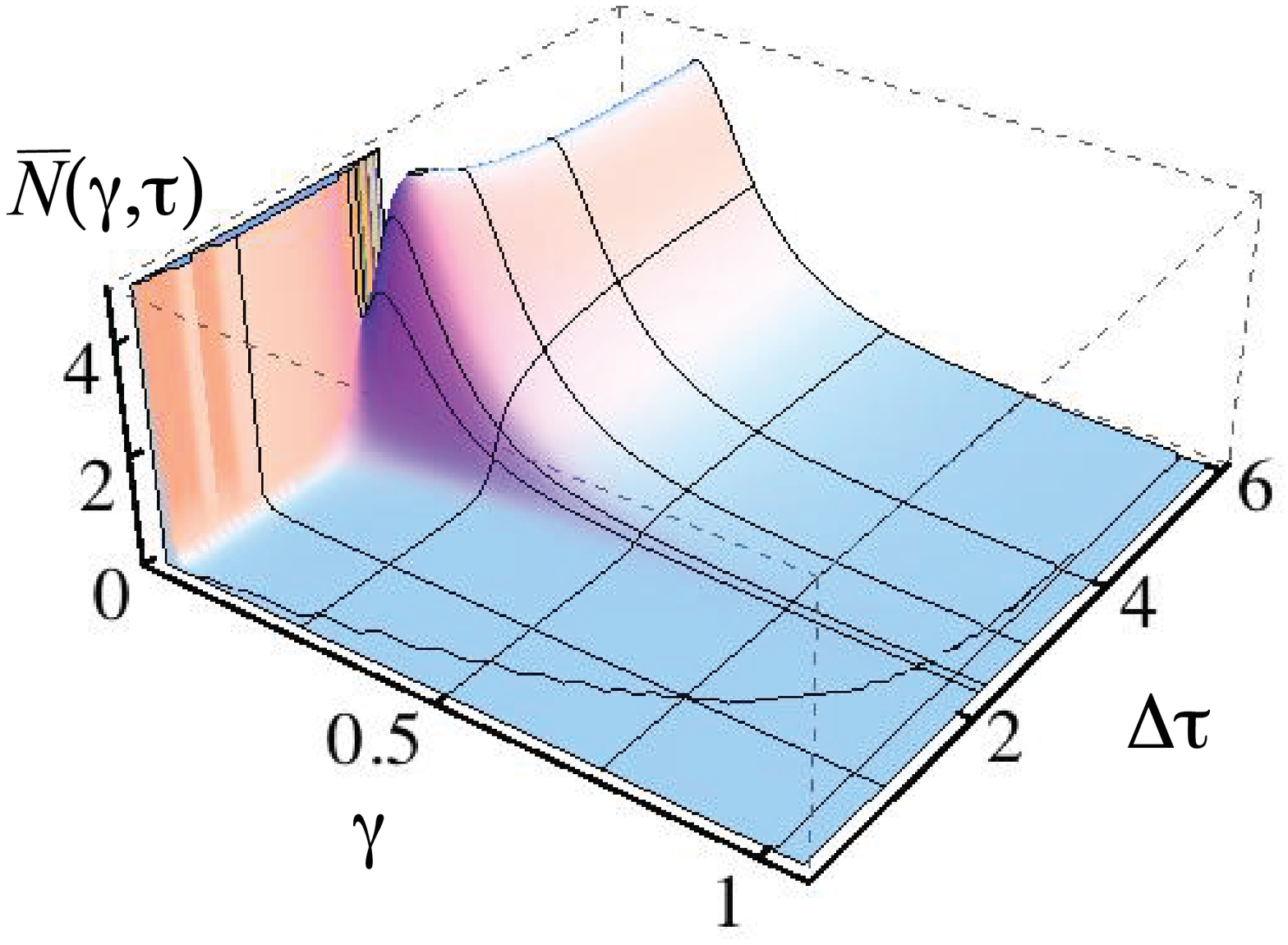}
\begin{center}
\caption{\label{fig:m01Nt3D}Time evolution of the distribution $\bar{N}(\gamma,\tau)$ for $n=1$ (top left), $n=2$ (top right), and $n=3$ (bottom).$m=0$. $t_r = 0.75$, ${\cal V}_0 = 1$. The lines at constant $\Delta\tau=\tau - \tau_0$ are depicted in Fig.~\ref{fig:m01Ntcuts} left column. The curved line in the $(\gamma,\Delta \tau)$ plane shows the maximal grain size as a function of time and corresponds to the curves of Fig.~\ref{fig:gammatau}.}
\end{center}
\end{figure}

Figure \ref{fig:m01Nt3D} displays the time-evolution of the distribution in three-dimensional plots for the various values of $m$ and $n$.

\begin{figure}[h]
\includegraphics[width=0.35\textwidth]{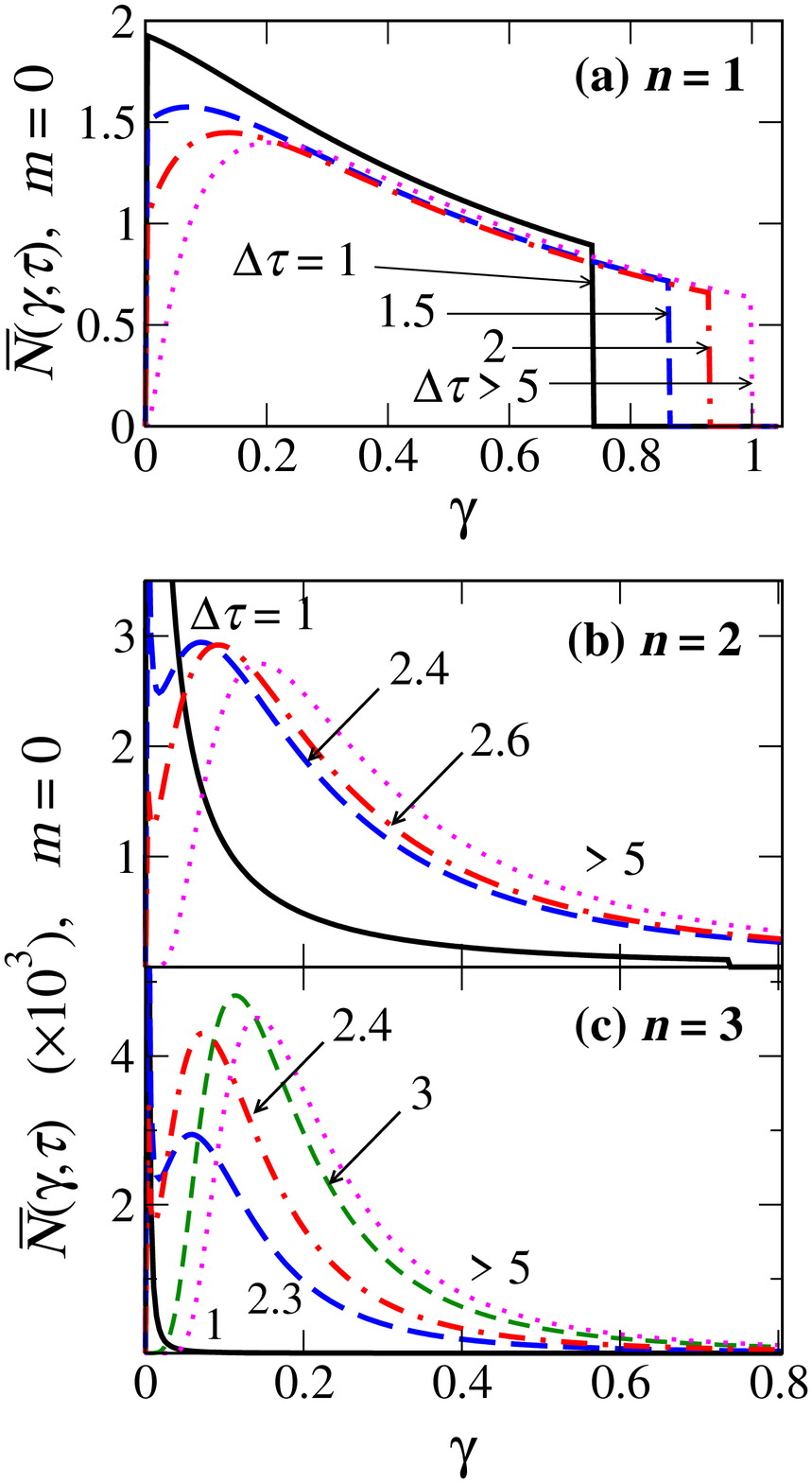}
\hspace*{0.2cm}
\includegraphics[width=0.335\textwidth]{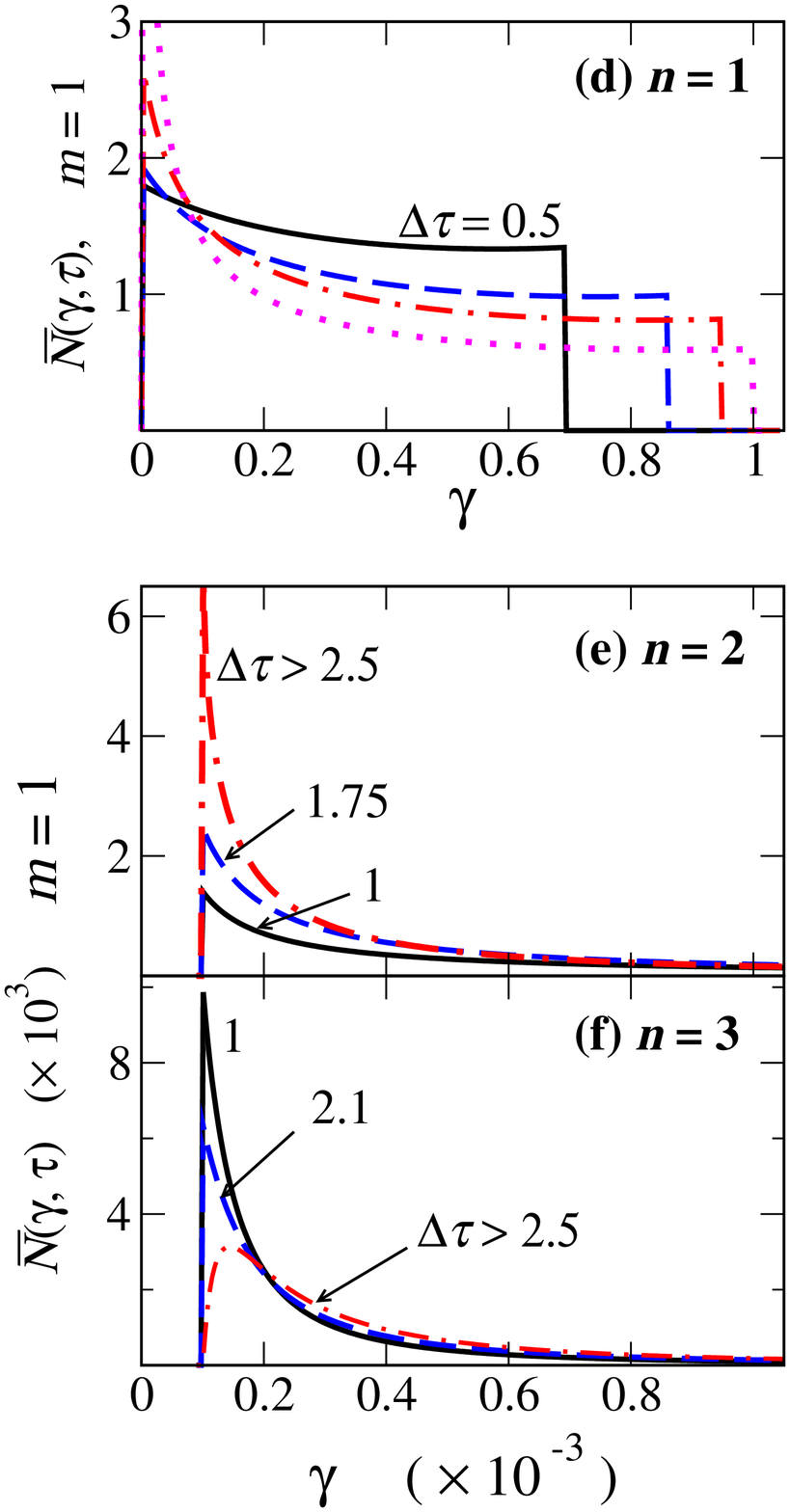}
\begin{center}
\caption{\label{fig:m01Ntcuts} $\bar{N}(\gamma,\tau)$ at different times $\Delta\tau\equiv \tau-\tau_0$ (cuts from Fig.~\ref{fig:m01Nt3D}). Left column: $m=0$, right column: $m=1$. Top row is $n=1$ for (a) and (d), $n=2$ for (b) and (e), and $n=3$ for (c) and (f). Note the abscissa and ordinate scales, especially for Figs.~(e) and (f). The time dependence is discussed in the text. $t_r = 0.75$, ${\cal V}_0 = 1$. The curves displayed for $m=0$ (left column) correspond to the thin lines at constant $\Delta\tau$ in Fig.~\ref{fig:m01Nt3D}.}
\end{center}
\end{figure}

Three observations can be made on Fig.~\ref{fig:m01Nt3D}. First, considering the intermediate to asymptotic time development when growth dominates over nucleation an increase in the value of $n$ leads to a more pronounced peak and decreasing value of the mean.  Thus, in accordance with the description of Figs.~\ref{fig:Nm0n} and \ref{fig:m0nVv0}, the faster the decay of the nucleation rate the smaller are the grains in average once crystallization is complete. Second, the specific time-dependence of the nucleation and growth rates has a notable influence on the early stage of crystallization. Whereas for $n=1$  (slow decay of the nucleation rate) the distribution has one broad peak at low $\gamma$ that further broadens and decreases in magnitude over time, for $n=2$ a sharp peak is observed at small radii of the grains (henceforth called the "nucleation peak") and a second smoother peak at larger values of $\gamma$ (called the "growth peak"). The case $n=3$ is similar, but the two peaks are better resolved. This is discussed in Fig.~\ref{fig:m01Ntcuts}. Figure \ref{fig:m01Nt3D} also displays a line delimiting the distribution at large $\gamma$. This line corresponds to $\gamma_m^{\max}(\tau)$ and was depicted in Fig.~\ref{fig:gammatau}. Note that the line saturates as $\tau\to\infty$ because of the time dependent growth rate. This contrasts with the result found in previous work, where $\gamma^{\max}(\tau\to\infty)\to\infty$.

Figure \ref{fig:m01Ntcuts} displays the cuts shown in Fig.~\ref{fig:m01Nt3D} for specific times $\tau$. The case $(n,m)=(1,0)$ (exponential and Gaussian decay of the growth and nucleation rates, respectively), Figure \ref{fig:m01Ntcuts}(a), has the shape of a lognormal distribution at infinite time. In this case, the grain size distribution has little structure at early stages. The case $m=1=n$ [Eq.~\eqref{GenSoltc}] is shown in Fig.~\ref{fig:m01Ntcuts}(d) and corresponds essentially to the product of two Gaussian functions with different prefactors that lead to the competition between a Gaussian time-decay of nucleation and a Gaussian saturation of the grain growth. Figure \ref{fig:m01Ntcuts}(d) displays the tail of the Gaussians.

More interesting are the cases $n = 2,3$. In Figs.~\ref{fig:m01Ntcuts}(b) and \ref{fig:m01Ntcuts}(c) ($m=0$) the sharp nucleation peak near $\gamma_c$ (typically $\gamma_c \sim 10^{-3}-10^{-4}$) and the broad growth peak described in the three-dimensional plots are clearly resolved. The time-evolution of these two peaks can be interpreted as follows. At early stages nucleation dominates the crystallization process leading to the formation of a large number of nuclei in a short time, and thus to a sharp peak at small radii. As time passes the nucleation peak is depleted and extends to larger radii [case $\Delta\tau = 1$ on Fig.~\ref{fig:m01Ntcuts}(b)]. Because the decay of nucleation is super Gaussian an intermediate-radii peak emerges as a result of the slower growth decay. The increase in the growth peak at intermediate values of $\gamma$ occurs at the expense of the nucleation peak (near $\gamma_c$). The front of the distribution shifts substantially as the broad peak starts emerging, then moves a little as the nucleation peak decreases, and finally shifts again as only growth determines the distribution.

\begin{figure}[h]
\includegraphics[width=0.4\textwidth,clip=true]{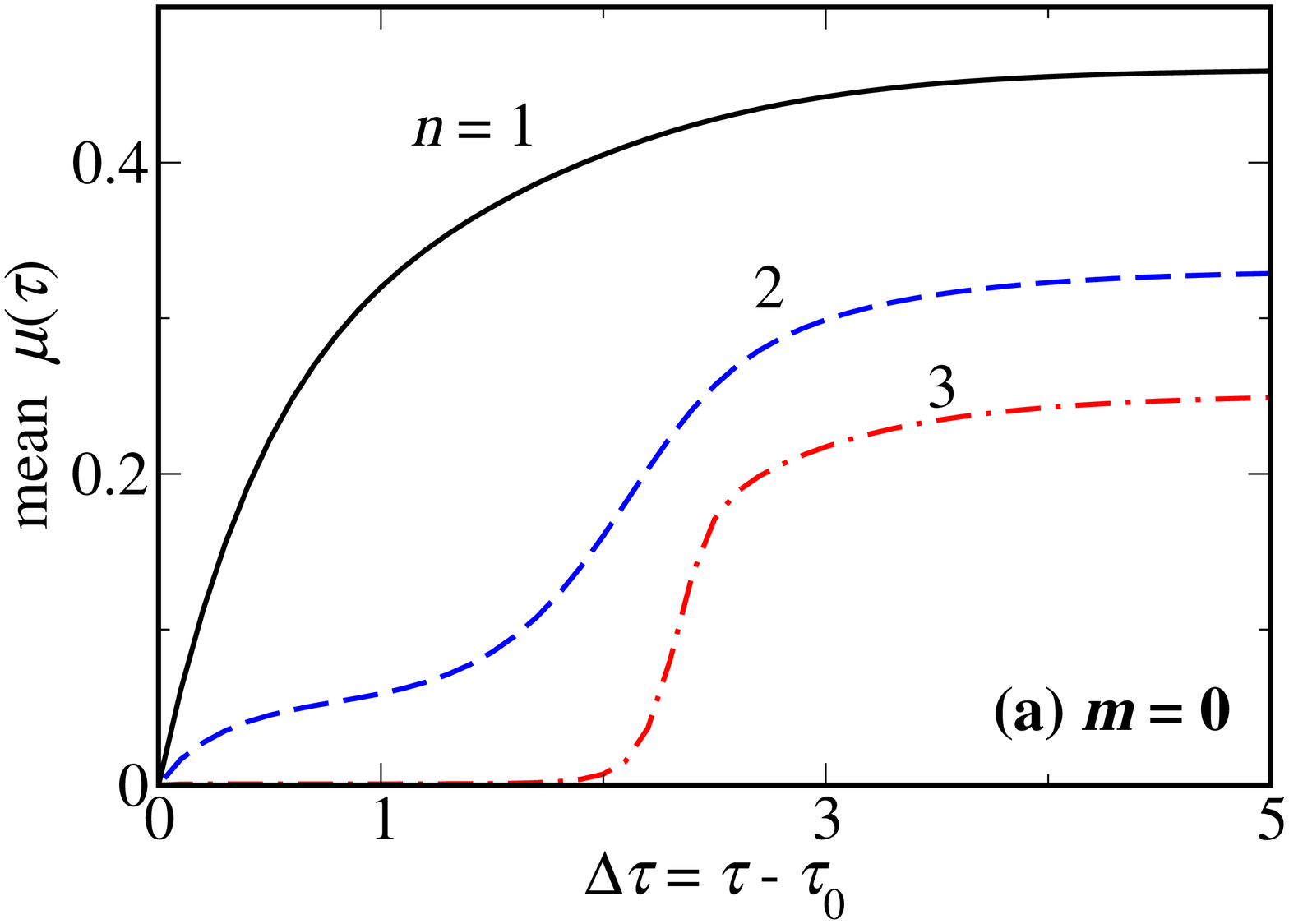}
\hspace*{0.2cm}
\includegraphics[width=0.4\textwidth,clip=true]{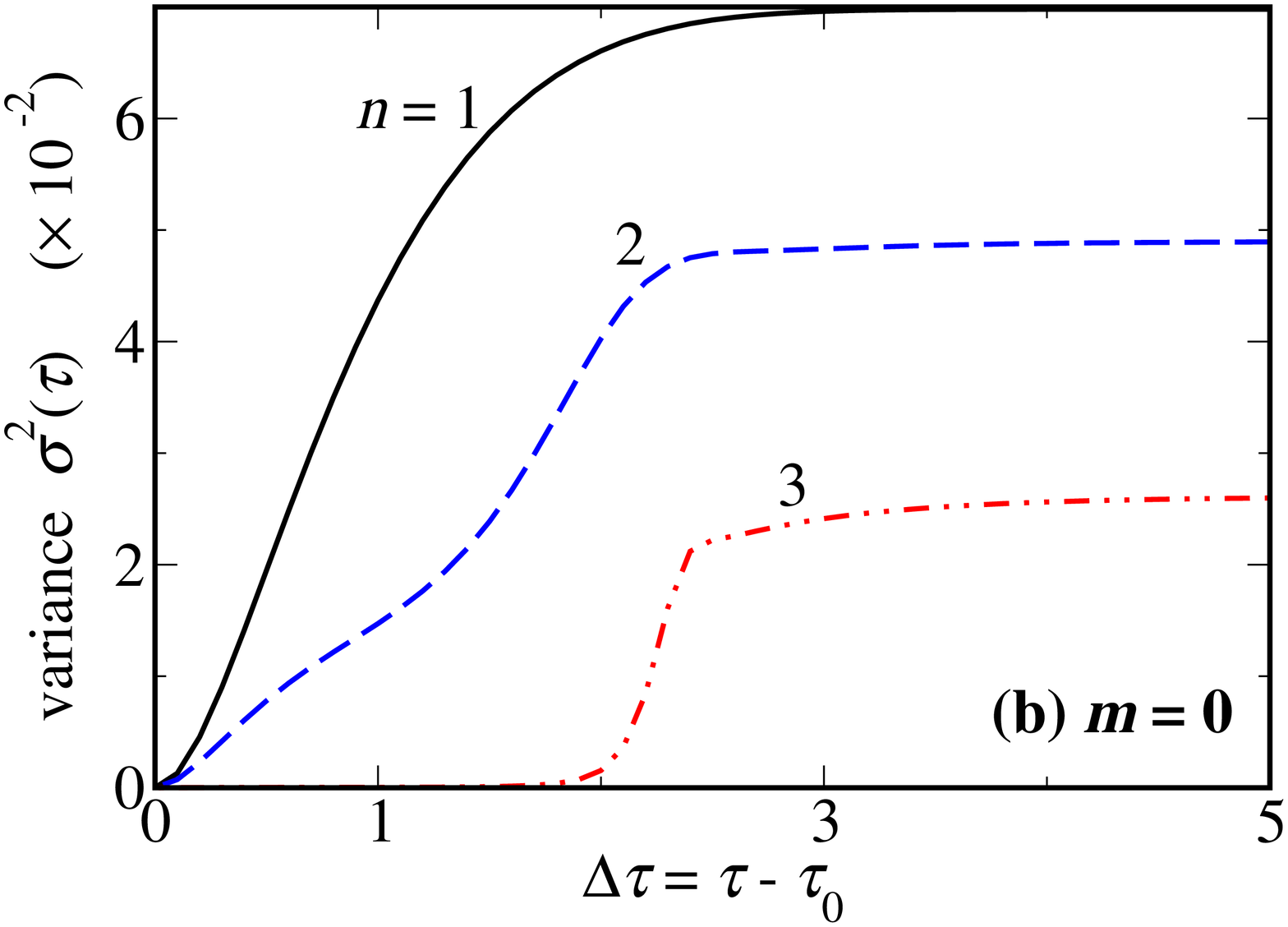}
\unitlength1cm
\caption{\label{fig:mean} Time-dependence of the mean $\mu(\tau)$, Eq.~\eqref{mom1}, and variance $\sigma^2(\tau)$, Eq.~\eqref{mom2}, of the grain size distribution. Both figures are for $m=0$, and $n=1$ (solid black), $2$ (blue dashed), and $3$ (red dash-dotted line). The mean and variance are those for the time-evolution of the grain size distribution depicted in Fig.~\ref{fig:m01Ntcuts}, left column. $t_r = 0.75$, ${\cal V}_0 = 1$.}
\end{figure}
Figure \ref{fig:mean} displays the time dependence of the mean and the variance for the case $m=0$ and $n=1,2,3$ (left column in Fig.~\ref{fig:m01Ntcuts}). We note that for $n=1$ both the mean and the variance increase smoothly to reach the value at infinite time. On the other hand, the cases $n=2$ and $3$ display a "kinklike" structure in the range $\Delta\tau \in [1,3]$. This structure reflects the fact that the nucleation peak decreases and broadens at first but remains very sharp, until $\Delta \tau \approx 1$. The appearance of the growth peak leads to an abrupt and large increase in the mean and the variance, which proceeds until the nucleation peak has completely disappeared. Once this stage is reached, both the mean and the variance increase marginally to their values at full crystallization.

The experimental observation of two peaks in the GSD is difficult and may often not be possible. For example, in the solid-phase crystallization of amorphous silicon only one peak is resolved.\cite{bergmann08,billMRS09} There are two possible reasons for this. One is that the nucleation peak is rather sharp and may be below experimental accuracy. The other is that the two peaks strongly depend on the time-evolution of nucleation and growth rates, but also on the early stage dynamics of nucleation. In the present model, we described the source term by a Dirac delta distribution [Eq.~\eqref{Ddelta}], which is a rather crude approximation of real systems. Instead, the delta distribution should be replaced, for example, by a Gaussian of finite variance, which spreads the formation of nuclei over $\gamma$.\cite{shinucleation} This modification broadens the nucleation peak and is expected to fill the dip between the nucleation and growth peaks and may prevent the resolution of the two peaks.

It is interesting to note that a Gaussian depletion of the growth rate obtained for $m=1$ also leads to the disappearance of two well-resolved peaks. This is depicted in Figs.~\ref{fig:m01Ntcuts}(e) and \ref{fig:m01Ntcuts}(f). Nonetheless, the dynamics of crystallization is otherwise very different for $m=0$ and $m=1$. One essentially observes a tall peak at small values of $\gamma$, the magnitude of which decreases and broadens in time to become the infinite time distribution. But the maximum of the peak for $m=1$ is orders of magnitude larger than for $m=0$ [note the different abscissa and ordinate scale of (e) and (f)]. In addition, because of the more rapid decay of the growth rate, the second peak has its maximum at smaller radius and is consequently fully absorbed by the nucleation peak. This is also the reason why the peak maximum barely shifts with time.

Figures \ref{fig:m01Ntcuts} and \ref{fig:mean} convey the message that the time dependence of nucleation {\it and} growth rates is essential for the time-evolution of crystallization processes, and also leads to very different shapes of the final GSD at full crystallization. 
 
\subsection{Discussion of the general solution and the derivation of a lognormal distribution}\label{ss:remarks}

We discuss the results of the previous paragraphs in the context of work done on random nucleation and growth processes and more generally work involving lognormal distributions. We showed in Ref.~\onlinecite{bergmann08} and Eqs.~\eqref{Nm=0n=1tinfty}, \eqref{Nm=0ntinftyrc}, and \eqref{Nlognormal} that a lognormal distribution is found at infinite time (more precisely, when $t\gg t_{cI},\,t_{cv}$.) This distribution has been so widely studied for many years that it may seem surprising that it has not been previously derived from a partial differential equation. To the best of our knowledge, this is, however, the case and is certainly true in the context of random nucleation and growth processes. Although observed in many instances, there are only few recorded attempts to actually derive this distribution from first principles. This is not to say that the lognormal distribution has never been derived.

In fact, studies involving the lognormal distribution can be divided in two categories. One category, which includes the majority of publications on the subject, {\it postulates} a lognormal distribution (or more involved varieties thereof).\cite{aitchinson69,kurtz80,crow88,williams91,schmelzer05} A fit to experimental data is then performed to determine the parameters of the distribution. Work belonging to this category does not derive the lognormal distribution and can, for example, not express the mean and variance in terms of fundamental parameters of the system.

The second category forms a much smaller group of publications, in which a derivation of the lognormal distribution is proposed. In an insufficiently cited paper,\cite{kolmogorov41} A.N.~Kolmogorov presented what is likely the first derivation of the lognormal distribution, applying probabilistic and statistical arguments to crushed powders, where previous experimental studies had shown that the lognormal distribution delivers a good fit to the data. Shortly thereafter, Epstein discussed the same grinding process and presented an alternative, somewhat clearer derivation,\cite{epstein47} demonstrating that under fairly general conditions the lognormal distribution is obtained as a result of the central limit theorem (see also Refs.~\onlinecite{kurtz80} and \onlinecite{redner90}). It is nowadays widely spread in the literature that a lognormal distribution can be obtained for multiplicative random processes (as opposed to additive processes, such as the random walk) using the central limit theorem applied to the natural logarithm of the product of probabilities. The limitations and problems of this approach  have been discussed in detail in Ref.~\onlinecite{redner90}, together with their remedies.

Our derivation differs from the work of Kolmogorov and Eptsein in several ways. First, our study of RNG processes starts at the opposite end of particle formation. Instead of breaking larger particles into smaller pieces, RNG starts with the creation of nuclei that grow over time. Thus, our description of the grain size distribution must include two physical phenomena: a source term accounting for the initial formation of grains and the growth of these grains. References \onlinecite{kolmogorov41} and \onlinecite{epstein47} do not contain any information on the source term (except for a postulated initial distribution) and how it affects the time-evolution of the particle size distribution. Second, we derive the lognormal distribution from a partial differential equation rather than probabilistic arguments. This is thus done in the same spirit as Avrami's approach to the kinetics of phase transformation, although the latter only determined the fraction of transformed material and not the grain size distribution discussed here. We do not refer to the central limit theorem in the derivation and overcome thereby limitations related to this approach.\cite{redner90} On the other hand, we explicitly use the conservation of mass, which is implicit in Kolmogorov and Epstein's derivations. Finally, contrary to the latter authors, our derivation enables us to write an explicit closed analytical expression for the grain size distribution at all times, and not only in the asymptotic limit. This allows us to study the time-evolution of the grain size distribution observed in the crystallization process of amorphous Silicon (see Refs.~\onlinecite{bergmann08} and \onlinecite{billMRS09}) and other materials. We also stress the fact that our derivation is done for arbitrary dimension of the grain formation and the truncated lognormal distribution (together with the caveats mentioned in Sec.~\ref{ss:diffeqRNG} and Appendix \ref{a:solutioninrandt}) is only one of the solutions obtained, namely in the case $n=m+1=1$ and in the asymptotic limit of large times.\\

Another remark concerns the partial differential equation and the determination of its analytical solution. Partial differential equations similar to the one established in Ref.~\onlinecite{bergmann08} for RNG processes and discussed here have been studied at length in the literature.\cite{PDEbooks} They have been applied to describe various phenomena in nature and, in some cases, an analytical solution has been derived.\cite{brock72,gelbard79,sekimoto,williams91} However, none of the papers have recognized and proven that the lognormal distribution is actually a solution of such a PDE. Our derivation clarifies why this has been overseen. Although a formal solution of the PDE can be obtained by quadrature, specific additional conditions on the functional form of $\mathbf{v}(\mathbf{r},t)$ and ${\cal D}(\mathbf{r},t)$ are required to derive the lognormal solution. Generally, when specified in the literature, polynomial forms in $r$ and simple exponentials in time have been considered for these functionals, and these do not lead to a lognormal-type distribution.\cite{brock72,gelbard79,williams91} The class of PDE considered here involves more complicated exponential forms of the time dependence that result from the physical analysis of RNG processes\cite{kolmogorov37,avrami39,mehljohnson} and these do lead to the obtained lognormal-type forms. In this context, it is interesting to note that the closest expression to a lognormal distribution is obtained for $d=1$, where the functional form of $\mathbf{v}(\mathbf{r},t)$ is exponential in time ($m=0$ and $\mathbf{v}$ has no dependence on $\mathbf{r}$), whereas ${\cal D}(\mathbf{r},t)$ is Gaussian in time ($n=1$). This leads to the conclusion that when a lognormal distribution is experimentally observed for grains described in terms of a scalar such as the volume or the average radius of the grains it is indicative not only of a particular size dependence of the source-and-growth terms but also of the dynamics of crystallization. The general conditions under which a lognormal distribution is a solution of the PDE is a topic of its own and will not be discussed further here.

Finally, the results obtained in this paper are not only interesting in themselves, but allow also to go one step further in the phenomenological description of RNG processes that may be useful for applications, since we determine all features of the distribution in terms of the fundamental parameters of the RNG model, namely, $I_0$, $v_0$, $t_{cv}/t_{cI}$, $t_0$, and $\rho_c$. The theory may, in principle, even be applicable to other phenomena not limited to physics and where distributions display, for example, a lognormal-like behavior asymptotically. As mentioned earlier, it is well known that such distributions are observed in biology, economics, sociology, etc.\cite{aitchinson69,crow88} Clearly, not all these phenomena can necessarily be described in terms of the differential equation \eqref{PDEgeneral}. But, it cannot be excluded that for certain problems, where a conservation law is present, and $\mathbf{v}_m(\mathbf{r},t)$ and ${\cal D}(\mathbf{r},t)$ have appropriate context-given meaning, the equation and solutions discussed here may be relevant. It is beyond the scope of this paper to discuss such generalization.

\section{Conclusions}\label{s:conclusions}

The paper presents a description of the time-evolution of the GSD in the crystallization of a $d-$dimensional solid. We provide an analytical derivation of the general solution of the partial differential equation for RNG and the Kolmogorov-Avrami-Mehl-Johson model for the fraction of material available for further crystallization. The general analytical expression for the GSD, Eq.~\eqref{GenSol1} with Eqs.~\eqref{eqsigmami} and \eqref{rates} (see also Appendix \ref{a:solutioninrandt} for expressions that can be used for experimental data analysis), has been divided into classes defined by $(d,n,m)$, where $d$ is the dimensionality of the growth process, $n$ and $m$ determine the time-decay rates of nucleation and growth, respectively. A key ingredient of the present theory is the introduction of an effective time-dependent growth rate in addition to the usual effective time-dependent nucleation rate. This new time dependence takes into account the fact that impingement inhibits the growth of grains, which reduces the actual average growth rate in time.

The main conclusions can be summarized as follows. First, the time decay of nucleation and growth rates plays a major role for the evolution of the distribution during crystallization. The time decay strongly affects the shape of the distribution, even at full crystallization. Thus, the time path followed by the system to reach the equilibrium distribution is essential.

Second, in the particular case of an exponential decay in time of the effective growth rate ($m=0$), the grain size distribution develops asymptotically in time to a form that is very close ($n=2,3$) or even essentially identical (for $n=1$) to the well-known lognormal distribution. The only difference being the presence of cutoffs at small and large grain sizes and an overall constant prefactor  [see Eqs.~\eqref{Nm=0ntinftyrc} and \eqref{Nlognormal}]. This result was pointed out in Ref.~\onlinecite{bergmann08} for $d=3$. Its extension to any dimension, and its significance are discussed in the present paper. It is remarkable that the derived GSD has a shape that is closely related to the lognormal distribution. It would be of interest to define more precisely the conditions on the functional form of source-and-growth terms under which a lognormal-like distribution is a solution of the PDE. Note that our study does not include possible additional contributions to the crystallization process. For example, coarsening is explicitly discarded from our considerations, although this phenomenon is also known to lead to a lognormal grain size distribution in the asymptotic time regime. This, however, is known for a long time. The realization that coarsening is not necessary, as discussed here and in Refs.~\onlinecite{bergmann97,bergmann97b} and \onlinecite{kumomi99}, is a more recent insight on crystallization.

Third, the relative time decay of nucleation and growth rates affects drastically the mean and the variance of the distribution. For the model used here to describe nucleation [a Dirac delta distribution, Eq.~\eqref{Ddelta}], the model even leads to the presence of two peaks at intermediate stages of crystallization. As discussed in Sec.~\ref{ss:tevol} the dip separating the two peaks is likely to disappear when replacing the Dirac distribution by a Gaussian nucleation rate with finite variance. This requires further work.

Finally, the presence of a time dependent effective growth rate leads to a finite maximal grain size at all times, including for $t\to\infty$ [Eqs.~\eqref{gammax} or Eq.~\eqref{gammaxrt}]. This contrasts with existing theories that have a constant effective growth rate and for which the distribution does not have a maximal grain size above which the distribution vanishes.

One interesting prediction of the model is the dependence of the distribution on the dimensionality of the crystallization process. For example, it is of interest to study the grain size distribution of thin films with various thicknesses. For film thicknesses smaller than the typical grain size, the experimental distribution should be best described with $d=2$. As the thickness of the film increases beyond the average grain size, the theoretical expression with $d=3$ should offer a better description of the data. Also, our theory provides a fundamental relation between parameters of the model and experimental quantities [Eq.~\eqref{paramrelation}].

An example of application of the theory to the time-evolution of the GSD in solid-phase crystallization of amorphous silicon has been presented in Refs.~\onlinecite{bergmann08} and \onlinecite{billMRS09}. We emphasize that contrary to many studies that rely on an {\it ad hoc} lognormal distribution to describe experimental data, we propose a description based on the creation and growth of grains. Thus, the shape of the distribution is derived and not postulated and relies on the knowledge of a few physical parameters describing the system. The derivation of the GSD has a predictive power for the physical description of the time-evolution\cite{bergmann08,billMRS09} that empirical approaches cannot offer.

Finally, the general formulation proposed in this paper for processes involving a source-and-sink term and a growth term should be applicable to a wider range of phenomena, where the grains are other entities, and where source-and-growth terms need to be defined appropriately. A few examples were offered in the introduction. We hope that the formulation and the derivation of the solution are written in general enough terms to initiate its application in other fields. We emphasize, however, that not all crystallization phenomena can be described by the KAMJ model. Therefore, we do not expect the solutions derived in this paper to be equally applicable to all materials. When they do, it is an indication that the Kolmogorov-Avrami-Mehl-Johnson model for the time-dependent fraction of the transformed phase is a good description of the crystallization process, as can be experimentally observed for the crystallization of amorphous silicon.\cite{bergmann96} When they do not, the original differential equation should be considered again, with other forms of the nucleation and growth rates, possibly including anisotropy and size dependence.

\acknowledgments{A.B.~and A.V.T.~are grateful to the Research Corporation and to SCAC at California State University Long Beach for the support. This paper has been finalized at BIAS in Bremen, Germany. A.B.~thanks the DAAD for making the visit possible, and members of BIAS for their hospitality.}
\newpage
\appendix
\section{Table of variables, parameters and relations}\label{a:parameters}

We provide a table (Table \ref{tab:definitions}) summarizing all variables, parameters and their definitions. In addition, we present below the table some relations between parameters.
\begin{longtable*}{@{\extracolsep{1in}}p{1.2in}p{4.8in}}
\caption{\label{tab:definitions}Definitions of variables, constants and functions used in the paper.}\\
$N(\mathbf{r},t)$, $\bar{N}(\gamma,\tau)$ & 
Grain size distribution (GSD), Eq.~\eqref{PDEgeneral}, and normalized (GSD) defined in Eq.~\eqref{Nnormalized}.\\
$\tilde{N}(\mathbf{r},t)$ & 
Auxiliary dimensionless function introduced in Eq.~\eqref{PDENtildedimless} to transform the partial differential equation (PDE) in a form that can be solved analytically.\\
$d$ & 
Dimension of the crystallization process. Although the calculations are valid for any $d$ we consider specifically $d=1,2,3$.\\\hline
$\mathbf{r}=(r_1,\dots,r_d)$ & 
$d-$dimensional vector of magnitude $r$, the components of which are the semi-axes $r_1\geq r_2\geq \dots \geq r_d$ of the ellipsoid that models a particular grain.\\
$\rho$ & 
Radius of the spherical grain. The latter is the limiting case of an ellipsoid with all semi-axes $r_j = \rho$ ($j=1,\dots,d$) and the previous $d-$dimensional vector reads $\mathbf{r} = (\rho,\dots,\rho)$. Consequently, $r = \sqrt{d}\, \rho$.\\
$\rho_c$ & 
Critical radius. Radius of the nucleus, the smallest grain that can be found in the system.\\
$\rho_m^{\max}(t)$ & Radius of the largest grain found in the sample at time $t$. The radius of the largest grain depends on the growth rate. Accordingly, the maximal grain size has an index $m$, which speficies which growth rate is considered, Eq.~\eqref{vt}.\\
$\rho_m^\infty$ & 
Radius of the largest grain found in the sample at $t\to\infty$ (full crystallization).\\
$\gamma = r/r_m^\infty = \rho/\rho_m^\infty$ & 
Dimensionless variable for the size of the grain. Except for Appendix \ref{a:solutioninrandt} all calculations are done using this dimensionless variable. In the text we omit the index $m$ for clarity.\\\hline
$t$, $t_0$ & 
Time variable $t$ and incubation time $t_0$. \\
$t_{cv}$, $t_{cI}$, $t_r$ & 
critical time of decay for the effective growth and nucleation rate, $v(t)$ and $I(t)$ in Eq.~\eqref{rates}. $t_r = \sqrt{t_{cv}/t_{cI}}$.\\
$\tau = t/\sqrt{t_{cv}t_{cI}}$ & 
Dimensionless time variable. $\tau_0$ is the corresponding dimensionless incubation time.\\\hline
$n$ , $m$ & 
Index of nucleation and growth laws, Eq.~\eqref{rates}, respectively. In the paper we focus on $n=d$ and $m=0,1$.\\
$I_0$, $v_0$ & 
Constant microscopic nucleation and growth rates.\\
${\cal I}_0$ , ${\cal V}_0$& 
dimensionless constants defined in Eq.~\eqref{dimlesscte}.\\
$Y_n(t)$, $Y_n^I(t)$, $Y_m^v(t)$ & 
KAMJ fraction of available space for crystallization and functions determining the time dependence of the nucleation and growth rates $I(t)$ and $v(t)$, respectively. See Eqs.~\eqref{KAMJ} and \eqref{rates}.\\
$I(t)$, $v(t)$ & 
Time dependent nucleation and growth rates, Eq.~\eqref{rates}.\\\hline
$\Omega$ & 
Volume of an ellipsoidal grain defined in Eq.~\eqref{Omegad}\\
$A_{c,d}$, $A_{\infty,d}$ & 
Surface of the nucleus and the largest grain at full crystallization, defined near Eq.~\eqref{PDEradius} and \eqref{PDENtildedimless}, respectively.\\\hline
$C_d$ & Constant prefactor of the grain size distribution, defined in Eq.~\eqref{Cd}.
\\
$\sigma_{m,i}(\gamma,\tau)$ & 
Function appearing in Eq.~\eqref{GenSol} and defined through Eq.~\eqref{eqsigmami}.\\
$\mu(\tau)$, $\sigma^2(\tau)$,  $\gamma_1(\tau)$ & 
mean, variance and skewness of the grain size distribution. They are defined in Eq.~\eqref{moments}. Note that $\gamma_1$ cannot be confused with the dimensionless size variable $\gamma$ which is never written with the index $m$ (see definition above.)
\end{longtable*}
A word of caution is necessary with respect to the definition of the rates $I_0$ and $v_0$. In expressions where quantities with dimensions (as in Appendix \ref{a:solutioninrandt}) are used, the definition of $I_0$ and $v_0$ depends on which variable is used to characterize the size of the grains $r$, $\rho$, or the diameter $g=2\rho$, as in Ref.~\onlinecite{bergmann08}. For example, the growth rate $v(t)$ appearing in terms of the different variables is related by $v^r = dr/dt = \sqrt{d}\,d\rho/dt = \sqrt{d}\,v^\rho$ and $v(\tau) = d\gamma/d\tau$ is given by Eq.~\eqref{vdimless}. Throughout the paper we omit the index $r,\rho$ or $g$ for the rates as the expression to use is univocally determined by the context. A similar argument applied to $I_0$ gives $I_0^r = I_0^\rho$. It turns out that with this precaution in mind, the expressions derived in the paper are essentially invariant with respect to the choice of variable, except the terms in the Heaviside function. This remark about the choice of variables is only relevant for Appendix \ref{a:solutioninrandt} below since the rest of the paper is written in dimensionless quantities and involves only $t_r$, ${\cal V}_0$, and ${\cal I}_0$ only.

\section{Main results written in quantities with physical units}\label{a:solutioninrandt}

The distributions derived in the main part of the paper were presented and analyzed using dimensionless quantities, $\gamma = r/r_m^\infty = \rho/\rho_m^\infty$ for the grain size and $\tau = t/\sqrt{t_{cv}t_{cI}}$ for time. Since one of the goals of the paper is to provide expressions for the grain size distribution that can be applied to the analysis of experimental data, we provide in this section the main results in quantities with physical units, using $\rho$ for the radius of the spherical grain and $t$ for the time. The expressions below take the exact same form when replacing the radius $\rho$ by the grain diameter $g$, under the condition that appropriate definition of $v_0$ and $I_0$ are considered (see discussion in the previous appendix).

According to Eq.~\eqref{GenSol} or Eq.~\eqref{GenSol1}, $N(\rho,t)$ is defined in units of number of grains per unit volume, per unit length (of the radius of the spherical grain) or, equivalently, $N\,d\rho$ has the units number of grains per unit volume. Thus, $I_0$ is expressed in number of grains per unit length and unit time while $v_0$ is in length per unit time.

For arbitrary $n$ but $m=0$, Eqs.~\eqref{Nm=0n} read as
\begin{subequations}\label{Nm=0nrt}
\begin{eqnarray}\label{Nm=0nrta}
N(\rho,t) &=& C_d \, \left(\frac{\rho_c}{\rho}\right)^{d-1}\,
\frac{\exp\left\{(-1)^n\left[t_r^2\ln\alpha_0\right]^{n+1}\right\}}{\alpha_0}\, \nonumber\\
&&\times
\Big[\Theta\left(\frac{\rho - \rho_c}{\rho_0^\infty-\rho_c}\right) - \Theta\left(\frac{\rho - \rho_0^{\max}(t)}{\rho_0^\infty-\rho_c}\right)\Big],
\end{eqnarray}
with $C_d$ defined in Eq.~\eqref{Cd},
\begin{eqnarray}\label{Nm=0nrtb}
\alpha_0(\rho,t) &=& \frac{\rho-\rho_c}{\rho_0^\infty-\rho_c}+e^{-(t-t_0)/t_{cv}},
\end{eqnarray}
\end{subequations}
and
\begin{eqnarray}\label{gammaxm0rt}
\rho_0^{\max}(t) &=& \rho_c + t_{cv}v_0 \, \left(1- e^{-(t-t_0)/t_{cv}}\right) \stackrel{t\to\infty}{\longrightarrow} \rho_c + t_{cv}v_0 \equiv \rho_0^\infty.
\end{eqnarray}
The latter is the maximal grain size found at time $t$ during crystallization and at $t\to \infty$ for $m=0$ and arbitrary $n$. Note that one often has $\rho_0^\infty \gg \rho_c$.

For infinite time (full crystallization), the above expression simplifies to
\begin{eqnarray}\label{Nm=0ntinftyrt}
N(\rho,t\to\infty) &=& C_d\, \left(\frac{\rho_c}{\rho}\right)^{d-1}\left(\frac{\rho_0^\infty-\rho_c}{\rho-\rho_c}\right)\,
\exp\left\{(-1)^n\left[t_r^2\ln\left(\frac{\rho-\rho_c}{\rho_0^\infty-\rho_c}\right)\right]^{n+1}\right\} \nonumber\\
&& \times \Big[\Theta\left(\frac{\rho-\rho_c}{\rho_0^\infty-\rho_c}\right) - \Theta\left(\frac{\rho-\rho_0^{\infty}}{\rho_0^\infty-\rho_c}\right)\Big].
\end{eqnarray}
Choosing $n=d=1$ ($m=0$) and $t\to\infty$, one obtains the expression derived in Ref.~\onlinecite{bergmann08}
\begin{eqnarray}\label{Nm=0n=1tinftyrt}
N(\rho,t\to\infty) &=& C_d\, \left(\frac{\rho_0^\infty-\rho_c}{\rho-\rho_c}\right)\,
\exp\left\{(-1)^n\left[t_r^2\ln\left(\frac{\rho-\rho_c}{\rho_0^\infty-\rho_c}\right)\right]^2\right\} \nonumber\\
&& \times
\Big[\Theta\left(\frac{\rho-\rho_c}{\rho_0^\infty-\rho_c}\right) - \Theta\left(\frac{\rho-\rho_0^{\infty}}{\rho_0^\infty-\rho_c}\right)\Big].
\end{eqnarray}
As discussed in the main text, this distribution is of the lognormal type as seen most conveniently by considering the case $\rho_c \ll \rho \leq \rho_0^\infty$
\begin{eqnarray}
\label{Nm=0n=1tinftyrt1}
N(\rho\gg \rho_c ,t\to\infty) &\approx& C_d\, \left(\frac{\rho_0^\infty}{\rho}\right)\,
\exp\left\{(-1)^n\left[t_r^2\ln\left(\frac{\rho}{\rho_0^\infty}\right)\right]^2\right\} \times
\left[\Theta\left(\frac{\rho}{\rho_0^\infty}\right) - \Theta\left(\frac{\rho}{\rho_0^\infty}-1\right)\right].
\end{eqnarray}
Defining the constants
\begin{eqnarray}
s = \left(\sqrt{2}t_r^2\right)^{-1},\quad M=\ln \rho_0^\infty, \quad C_{\log} = \frac{I_0}{v_0}\,\frac{\sqrt{\pi}}{2\,t_r^2} \rho_0^\infty,
\end{eqnarray}
one can write for $\rho\in (0,\rho_0^\infty)$
\begin{eqnarray}\label{Nlognormal}
N(\rho\gg\rho_c,t\to\infty) &\approx& C_{\log} \frac{1}{\sqrt{2\pi}\,s\,\rho}\,\exp\left\{-\frac{\left(\ln\rho - M\right)^2}{2s^2} \right\} = C_{\log} f_{\log}(\rho),
\end{eqnarray}
where $f_{\log}(\rho)$ is the lognormal distribution with $M$ as the mean and $s$ as the standard deviation of the variable's logarithm. Note that the range of finite values of the distribution [ Eq.~\eqref{Nlognormal}] is $(0,\rho_0^\infty)$, while for the standard lognormal distribution it is $[0,\infty)$. This physically justified limited range of finite values, together with the presence of the multiplicative constant $C_{\log}$ in front of the expression lead to term the derived distribution as being lognormal-like.\\

Consider now the case $m=1$. For arbitrary $n$ [Eqs.~\eqref{Nm=1n}], we have
\begin{subequations}\label{Nm=1nrt}
\begin{eqnarray}\label{Nm=1nart}
N(\rho,t) &=& C_d\, \left(\frac{\rho_1^\infty}{\rho}\right)^{d-1}  \exp\left[\left(\mbox{erf}^{-1}\alpha_1\right)^2-\left(t_r^2\,\mbox{erf}^{-1}\alpha_1\right)^{n+1}\right]  \nonumber\\
&&\times \left\{  \Theta\left(\frac{\rho-\rho_c}{\rho_1^\infty - \rho_c}\right) - \Theta\left(\frac{\rho-\rho_1^{\max}(t)}{\rho_1^\infty - \rho_c}\right)\right\},
\end{eqnarray}
with
\begin{eqnarray}\label{Nm=1nbrt}
\alpha_1(\rho,t)  = {\rm erf}\left(\frac{t-t_0}{t_{cv}}\right) - \frac{\rho-\rho_c}{\rho_1^\infty - \rho_c}
\stackrel{t\to\infty}{\longrightarrow}
1 - \frac{\rho-\rho_c}{\rho_1^\infty - \rho_c}.
\end{eqnarray}
\end{subequations}
and
\begin{eqnarray}\label{gammaxm1rt}
\rho_1^{\max}(t) &=& \rho_c + t_{cv}v_0\, \frac{\sqrt{\pi}}{2}\,{\rm erf}\left(\frac{t-t_0}{t_{cv}}\right)
\stackrel{t\to\infty}{\longrightarrow} \rho_c + t_{cv}v_0\, \frac{\sqrt{\pi}}{2} \equiv \rho_1^\infty.
\end{eqnarray}
At infinite time, this expression reduces to a form that cannot be readily cast into a lognormal-type distribution, as discussed in the paper.

Finally, the general expression, Eq.~\eqref{GenSol1} for any $n$ and $m$ reads as
\begin{subequations}
\begin{eqnarray}\label{GenSolrt}
N(\rho,t) &=&
C_d\, \left(\frac{\rho_m^\infty}{\rho}\right)^{d-1} \exp\left[\left(\frac{\sigma_{m}-t_0}{t_{cv}}\right)^{m+1} -\left(\frac{\sigma_{m}-t_0}{t_{cI}}\right)^{n+1}\right] \nonumber\\
&&\times
\left\{  \Theta\left(\frac{\rho-\rho_c}{\rho_m^\infty - \rho_c}\right) - \Theta\left(\frac{\rho-\rho_m^{\max}(t)}{\rho_m^\infty - \rho_c}\right)\right\},
\end{eqnarray}
where $\sigma_{m}(\rho,t)$ is solution of
\begin{eqnarray}\label{eqsigmamirt}
\rho = \rho_c + u_m(\sigma_{m},t) = \rho_c + \int_{\sigma_m}^{t} v_m(t')\,dt'
\end{eqnarray}
with $v_m(t)$ defined in Eq.~\eqref{vt}, and $\rho_m^{\max}(t)$ is obtained in a similar way as Eq.~\eqref{gammax} 
\begin{eqnarray}\label{gammaxrt}
\rho_m^{\max}(t)
&=&\rho_c + u_m(t_0,t) \nonumber\\
&=& \rho_c + t_{cv}v_0 \left\{\Gamma\left[\frac{m+2}{m+1}\right] - \frac{1}{m+1}\Gamma\left[\frac{1}{m+1},\left(\frac{t-t_0}{t_{cv}}\right)^{m+1} \right]\right\}.
\end{eqnarray}
\end{subequations}
The maximal grain size $\rho_m^\infty = \lim_{\tau\to\infty}\rho_m^{\max}(\tau)$ obtained once crystallization is completed ($t\gg \max\{t_{cv}, t_{cI}\}$) immediately follows from the above expression
\begin{eqnarray}\label{ginf}
 \lim_{\tau\to\infty} \rho_m^{\max}(\tau) = \rho_m^\infty =
 \rho_c + \rho_m^\infty {\cal V}_0 \, \Gamma\left[\frac{m+2}{m+1}\right] =
 \rho_c + t_{cv} v_0 \, \Gamma\left[\frac{m+2}{m+1}\right].
\end{eqnarray}
For example,
\begin{subequations}\label{gammamaxatinfty}
\begin{eqnarray}
\rho^{\infty}_0 &=& \rho_c + t_{cv}v_0, \quad m=0,\\
\rho^{\infty}_1 &=& \rho_c +\frac{\sqrt{\pi}}{2}\,  t_{cv}v_0,\quad m=1.
\end{eqnarray}
\end{subequations}
This leads to an interesting relation between fundamental quantities of the model. Reformulating Eq.~\eqref{ginf}, one obtains  Eq.~\eqref{paramrelation} for any non-negative integer $m$. This relation quantity can, for example, be used to determine the critical time $t_{cv}$ once $\rho_m^\infty$ and $v_0$ have been measured, and taking into account the fact that, in general, the inequality $\rho_m^\infty\gg \rho_c$ holds.
One can also write Eq.~\eqref{ginf} in dimensionless quantities
\begin{eqnarray}\label{relationV0gamc}
1 = \gamma_c + {\cal V}_0 \,\Gamma\left[\frac{m+2}{m+1}\right].
\end{eqnarray}
This has been used to obtain Eq.~\eqref{gammax}.

To compare the above expressions with experimental data, it is best to normalize the distribution
\begin{subequations}\label{Nnormalizedrhot}
\begin{eqnarray}
\bar{N}(\rho,t) = \frac{N(\rho,t)}{N(t)},
\end{eqnarray}
where the total number of grains at time $t$ is given by
\begin{eqnarray}
N(t) = \int_0^\infty\,N(\rho,t) \,d\rho = \int_{\rho_c}^{\rho_m^{\max}(t)} N(\rho,t)\,d\rho.
\end{eqnarray}
\end{subequations}
This removes the coefficient $C_d$.

In Sec.~\ref{ss:moments} we defined the three first moments in terms of the dimensionless radius $\gamma = \rho/\rho_m^\infty$ for the normalized GSD. These quantities are correspondingly defined for GSD in terms of $\rho$ and we only write the relations between the two definitions:
\begin{subequations}\label{momentsrho}
\begin{eqnarray}
\mu_\rho(t) &=& \rho_m^\infty \mu,\\
\sigma_\rho^2(t) = \mu_{2,\rho}(t) &=& \left(\rho_m^\infty\right)^2\mu_{2},\\
\gamma_{1,\rho}(t) &=& \gamma_{1}.
\end{eqnarray}
\end{subequations}
The definitions in terms of $\rho$ are supplied by the index. Note that if one considers unnormalized quantities, these equations have to be modified accordingly.

\section{Solution of Eq.~\eqref{PDEradius} and proof of Eqs.~\eqref{prodHeavisidefinal}}\label{a:solution}

\subsection{Solution of Eq.~\eqref{PDEradius}}\label{a:solutionPDE}
We solve the partial differential equation \eqref{PDEradius} and derive Eqs.~\eqref{GenSol} and \eqref{GenSol1}. Using the Laplace transform,
\begin{eqnarray}
\tilde{N}(k,\tau) \equiv {\cal L}\left[\tilde{N}(\gamma,\tau)\right] = \int_0^\infty \tilde{N}(\gamma,\tau) e^{-k\gamma} d\gamma,
\end{eqnarray}
with $k\in \Bbb{R}$, Eq.~\eqref{PDEradius} transforms into a first-order ordinary differential equation in time
\begin{eqnarray}\label{diffeqLaplace}
\frac{\partial}{\partial \tau} \tilde{N}(k,\tau) + \frac{k}{t_r}\,v_m(\tau)\,\tilde{N}(k,\tau) = t_r\,I_n(\tau)e^{-k\gamma_c}.
\end{eqnarray}
where $v_m(\tau)$ and $I_n(\tau)$ are given by Eq.~\eqref{ratesdimless}. The solution of this equation can be immediately found to have the form
\begin{eqnarray}\label{Nktaugen}
\tilde{N}(k,\tau) = t_r\,e^{-k\gamma_c}\,\int_{\tau_0}^\tau I_n(\sigma)\,e^{- k\,u_m(\sigma,\tau)}\,d\sigma,
\end{eqnarray}
where we used the definition \eqref{umtau}.
The inverse Laplace transform of Eq.~\eqref{Nktaugen} leads to the following form of $N(\gamma,\tau)$:

\begin{subequations}\label{invLNgtau}
\begin{eqnarray}
\tilde{N}(\gamma,\tau) = t_r\, {\cal I}_0 \int_{\tau_0}^\tau\,d\sigma\,\,e^{-A_n(\sigma)} \,\,\delta\left(\gamma  - B_m(\sigma,\tau)\right),
\end{eqnarray}
with
\begin{eqnarray}
A_n(\sigma) = t_r^{n+1}\left(\sigma-\tau_0\right)^{n+1},\quad
B_m(\sigma,\tau) = \gamma_c + u_m(\sigma,\tau)
\end{eqnarray}
\end{subequations}
This result was obtained by exchanging the order of the integrals over $k$ and $\sigma$. This operation imposes convergence conditions on the integrand, which are satisfied for the KAMJ model for nucleation and growth.

We now use the relation
\begin{subequations}\label{deltaeqsi}
\begin{eqnarray}\label{delta}
\delta\left(\gamma  - B_m(\sigma,\tau)\right) = \sum_i \frac{\delta(\sigma-\sigma_{m,i})}{\left| \frac{\partial}{\partial \sigma} B_m(\sigma,\tau)\right|_{\sigma = \sigma_{m,i}}},
\end{eqnarray}
where we sum over all solutions $\sigma_{m,i}$ of the equation
\begin{eqnarray}\label{eqsi}
\gamma  = B_m(\sigma_{m,i},\tau),
\end{eqnarray}
\end{subequations}
with $\left. \frac{\partial}{\partial \sigma} B_m(\sigma,\tau)\right|_{\sigma_{m,i}} \neq 0$. This is Eq.~\eqref{eqsigmami}. Note that $\sigma_{m,i} = \sigma_{m,i}(\gamma, \tau)$. For the KAMJ model, the sum over $i$ reduces to one term because the function $B_m$ is monotonous in $\sigma$ for arbitrary values of $\tau$ and $m$. Furthermore, the condition on the derivative of $B_m$ is satisfied for $\sigma_{m,i} < \infty$, which is always true.

Taking into account the fact that Eqs.~\eqref{invLNgtau} are finite when $\sigma_{m,i} \in [\tau_0,\tau]$, we obtain the general solution of Eq.~\eqref{PDEradius} in the form
\begin{eqnarray}\label{NgtauB1}
\tilde{N}(\gamma ,\tau) &=& t_r\,{\cal I}_0 \sum_i \, \frac{e^{-A_n(\sigma_{m,i})}}{\left|\frac{\partial}{\partial \sigma} B_m(\sigma_{m,i},\tau)\right|} \Theta(\tau-\sigma_{m,i})\,\Theta(\sigma_{m,i}-\tau_0),
\end{eqnarray}
which for the KAMJ model becomes
\begin{eqnarray}\label{NgtauB2}
\tilde{N}(\gamma ,\tau) &=&
t_r^2\,\frac{{\cal I}_0}{{\cal V}_0} \sum_i \exp\left\{\left[\left(\frac{\sigma_{m,i}-\tau_0}{t_r}\right)^{m+1} - t_r^{n+1}\left(\sigma_{m,i}-\tau_0\right)^{n+1}\right]\right\}\,\Theta(\tau-\sigma_{m,i})\,\Theta(\sigma_{m,i}-\tau_0).
\end{eqnarray}
Equation \eqref{GenSol} follows from this expression. The expressions for $\sigma_{m,i}(\gamma,\tau)$ are determined from the solution of Eq.~\eqref{eqsi} and explicit expressions for $m=0,1$ were determined for the KAMJ model in Eqs.~\eqref{sigmam}.\\

\subsection{Proof of Eqs.~\eqref{prodHeavisidefinal}}\label{a:solutionHeaviside}

To prove that Eq.~\eqref{GenSol1} is equivalent to Eq.~\eqref{GenSol}, we have to prove Eq.~\eqref{prodHeavisidefinal}. 
The product of Heaviside functions on the left-hand side of Eq.~\eqref{prodHeavisidefinal} is equivalent to the statement $\tau \ge \sigma_{m,i} \ge \tau_0\ge 0$. Using Eq.~\eqref{umtau} and noting that if the expression for $v_m$ is non-negative [as is the case of Eq.~\eqref{vdimless}], the previous condition on $\sigma_{m,i}$ implies $u_m(\sigma_{m,i},\tau) \ge 0$. From Eq.~\eqref{eqsigmami}, we immediately conclude that $\gamma \ge \gamma_c$, which is known to be true by definition since $\gamma = \rho/\rho_m^\infty$.

On the other hand, for non-negative $v_m$ the inequality $\tau \geq \sigma_{m,i} \geq \tau_0\geq 0$ also implies $0\leq u_m(\sigma_{m,i},\tau) \leq u_m(\tau_0,\tau)$. Using again Eq.~\eqref{eqsigmami}, this condition can be written as $\gamma - \gamma_c \leq u_m(\tau_0,\tau)$ or, with Eq.~\eqref{gammaxeq}, $\gamma \leq \gamma_c+ u_m(\tau_0,\tau) = \gamma_m^{\max}(\tau)$. This proves Eq.~\eqref{prodHeavisidefinal} and, consequently, with Eq.~\eqref{NgtauB2}, completes the derivation of Eq.~\eqref{GenSol1}.


\end{document}